\documentclass{article}

\usepackage[
backend=biber,
style=numeric,
sorting=ynt
]{biblatex}

\usepackage{paralist}
\usepackage{graphicx} 
\providecommand{\keywords}[1]
{
  \small	
  \textbf{\textit{Keywords---}} #1
}
\usepackage{hyperref}
\usepackage{cleveref}
\usepackage{subfig}
\usepackage[dvipsnames]{xcolor}
\usepackage{listings}
\lstdefinelanguage{SPARQL}{
  basicstyle=\small\ttfamily,
  columns=fullflexible,
  breaklines=false,
  sensitive=true,
  frame=bt,
  aboveskip=1em,
  belowskip=1em,
  xleftmargin=.5em,
  xrightmargin=.5em,
  framexleftmargin=.5em,
  framextopmargin=.5em,
  framexbottommargin=.5em,
  framexrightmargin=.5em,
  tabsize = 2,
  showstringspaces=false,
  morecomment=[l][\color{gray}]{\#},       
  morecomment=[n][\color{blue}]{<http}{>}, 
  morestring=[b][\color{OliveGreen}]{\"},  
  keywordsprefix=?,
  classoffset=0,
  keywordstyle=\color{Sepia},
  morekeywords={},
  classoffset=1,
  keywordstyle=\color{Purple},
  morekeywords={rdf,rdfs,owl,xsd,purl},
  classoffset=2,
  keywordstyle=\color{MidnightBlue},
  morekeywords={
    SELECT,CONSTRUCT,DESCRIBE,ASK,WHERE,FROM,NAMED,PREFIX,BASE,OPTIONAL,
    FILTER,GRAPH,LIMIT,OFFSET,SERVICE,UNION,EXISTS,NOT,BINDINGS,MINUS,a,
    BIND,VALUES,CONTAINS
  }
}
\lstdefinelanguage{SQL}{
  basicstyle=\small\ttfamily,
  columns=fullflexible,
  breaklines=false,
  sensitive=true,
  frame=bt,
  aboveskip=1em,
  belowskip=1em,
  xleftmargin=.5em,
  xrightmargin=.5em,
  framexleftmargin=.5em,
  framextopmargin=.5em,
  framexbottommargin=.5em,
  framexrightmargin=.5em,
  tabsize = 2,
  showstringspaces=false,
  morecomment=[l][\color{gray}]{\#},       
  morecomment=[n][\color{blue}]{<http}{>}, 
  morestring=[b][\color{OliveGreen}]{\"},  
  keywordsprefix=?,
  classoffset=0,
  keywordstyle=\color{Sepia},
  morekeywords={},
  classoffset=1,
  keywordstyle=\color{Purple},
  morekeywords={rdf,rdfs,owl,xsd,purl},
  classoffset=2,
  keywordstyle=\color{MidnightBlue},
  morekeywords={
    SELECT,CONSTRUCT,DESCRIBE,ASK,WHERE,FROM,NAMED,PREFIX,BASE,OPTIONAL,
    FILTER,GRAPH,LIMIT,OFFSET,SERVICE,UNION,EXISTS,NOT,BINDINGS,MINUS,
    DISTINCT
  },
  classoffset=3,
  keywordstyle=\color{RedOrange},
  morekeywords={
    ST_ASTEXT,ST_BUFFER,ST_INTERSECTS,ST_TRANSFORM, GEOGRAPHY
  }
}
\usepackage{booktabs}
\usepackage{comment}

\usepackage{color,soul}
\usepackage{xspace}
\newcommand{\protege}{Prot\'{e}g\'{e}\xspace}

\title{Integrating 3D City Data through Knowledge Graphs}
\author{Linfang Ding$^1$ \and Guohui Xiao$^{2,3,4}$ \and Albulen Pano$^5$ \and Mattia Fumagalli$^5$ \and Dongsheng Chen$^7$ \and Yu Feng$^7$ \and Diego Calvanese$^{5,6,4}$ \and Hongchao Fan$^1$ \and Liqiu Meng$^7$}

\date{%
    $^1$Norwegian University of Science and Technology, 7491 Trondheim, Norway \\
    $^2$University of Bergen, 5007 Bergen, Norway \\
    $^3$University of Oslo, 0373 Olso, Norway \\
    $^4$Ontopic S.r.l., 39100 Bolzano, Italy \\
    $^5$Free University of Bozen-Bolzano, 39100 Bolzano, Italy \\
    $^6$Ume\aa{} University, 90187 Ume\aa{}, Sweden \\
    $^7$Technical University of Munich, 80333 Munich, Germany 
}

\begin{document}

\maketitle

\begin{abstract}    

CityGML is a widely adopted standard by the Open Geospatial Consortium (OGC) for representing and exchanging 3D city models. 
The representation of semantic and topological properties in CityGML makes it possible to query such 3D city data to perform analysis in various applications, e.g., security management and emergency response, energy consumption and estimation, and occupancy measurement.  
However, the potential of querying CityGML data has not been fully exploited. The official GML/XML encoding of CityGML is only intended as an exchange format but is not suitable for query answering. The most common way of dealing with CityGML data is to store them in the 3DCityDB system as relational tables and then query them with the standard SQL query language. 
%
%
Nevertheless, for end users, it remains a challenging task to formulate queries over 3DCityDB directly for their ad-hoc analytical tasks, because there is a gap between the conceptual semantics of CityGML and the relational schema adopted in 3DCityDB. 
In fact, the semantics of CityGML itself can be modeled as a suitable ontology.
The technology of Knowledge Graphs (KGs), where an ontology is at the core, is a good solution to bridge such a gap.
Moreover, embracing KGs makes it easier to integrate with other spatial data sources, e.g., OpenStreetMap and existing (Geo)KGs (e.g., Wikidata, DBPedia, and GeoNames), and to perform queries combining information from multiple data sources.
%
%
In this work, we describe a CityGML KG framework to populate the concepts in the CityGML ontology using declarative mappings to 3DCityDB, thus exposing the CityGML data therein as a KG. 
%
To demonstrate the feasibility of our approach, we use CityGML data from the city of Munich as test data and integrate OpenStreeMap data in the same area.
Finally, we collect real-world geospatial analytical tasks and show that they can be formulated as intuitive GeoSPARQL queries. We test three state-art-of-art KG systems, Ontop, Apache Jena, and GraphDB, and confirm that the queries can be evaluated efficiently over the generated KG.
\end{abstract}



\section{Introduction}

3D city data has been increasingly used to perform analysis in various applications, e.g., security management and emergency response, energy consumption and estimation, and occupancy measurement. A widely adopted standard by the Open Geospatial Consortium (OGC) for representing and exchanging 3D city models is \textit{CityGML}~\cite{citygml,ogc-citygml-v3}. It defines the three-dimensional geometry, topology, semantics, and appearance of the most relevant topographic objects in urban or regional contexts. 
The representation of semantic and topological properties in CityGML makes it possible to query such 3D city data to perform analysis.

At the implementation level, CityGML is defined as a GML application schema for the Geography Markup Language (GML)~\cite{citygml}. In its most common implementation, CityGML datasets consist of a set of XML files and possibly some accompanying image files that are used as textures. Each text file can represent a part of the dataset, such as a specific region, a specific type of object (such as a set of roads), or a predefined Level of Detail (LoD). The structure of a CityGML file is a hierarchy that ultimately reaches down to individual objects and their attributes. These objects have a geometry that is described using GML.
Another important implementation of CityGML is 3DCityDB~\cite{yao20183dcitydb}, which is a free 3D geo-database solution for CityGML-based 3D city models. 3DCityDB has been developed as an open source and platform-independent software suite to facilitate the development and deployment of 3D city model applications. The 3DCityDB software package consists of a database schema for spatially enhanced relational database management systems (Oracle Spatial or PostgreSQL/PostGIS) with a set of database procedures and software tools allowing to import, manage, analyze, visualize, and export virtual 3D city models according to the CityGML standard.

However, the potential of querying CityGML data has not been fully exploited. The official GML/XML encoding of CityGML is only intended as an exchange format but is not suitable for query answering. The most common way of dealing with CityGML data is to store them in the 3DCityDB system as relational tables and then query them with the standard SQL query language. 
%
%
Nevertheless, for end users, it remains a challenging task to formulate queries over 3DCityDB directly for their ad-hoc analytical tasks, because there is a gap between the conceptual semantics of CityGML and the relational schema adopted in 3DCityDB.

One possibility to bridge this gap is to use \emph{semantic technology}, which is concerned with the challenges posed by data with a complex structure and associated knowledge.
At the core of solutions based on semantic technology, we typically have an \emph{ontology} to provide semantics to the data. In computer science, the term “ontology” denotes a concrete artifact that conceptualizes a domain of interest and allows one to view the information and data relevant for that domain in a sharable and coherent way.
In the CityGML standard, the semantics is defined as a collection of UML diagrams, which can be naturally regarded as an ontology.
The instances of an ontology are \emph{knowledge graphs (KGs)}~\cite{DBLP:journals/csur/HoganBCdMGKGNNN21}, where data is structured in the form of a graph. Domain objects and data values are represented as nodes of such a graph, and properties of objects are encoded as edges.
For CityGML, the nodes in a KG could represent instances of buildings, streets, and surfaces, among others. 
Moreover, embracing KGs makes it possible to integrate with existing
KGs, e.g., Wikidata~\cite{wikidata}, DBPedia~\cite{dbpedia-swj}, \href{http://geonames.org/}{GeoNames}, and
LinkedGeoData~\cite{SLHA11,2021-jws-lgd}. This allows us to express interesting queries that require combining information from multiple sources.

In this work, we describe a CityGML KG framework to expose CityGML data as 
a Knowledge Graph and to integrate it with other data (e.g., OSM data). The CityGML KG or the integrated KG can be queried using the standard GeoSPARQL query language.
%
To demonstrate the feasibility of this framework, we use the 3D CityGML building data at LoD2 of the municipality of Munich, Germany as test data. We adopt and extend the CityGML ontology created by the University of Geneva and develop a suitable R2RML mapping to 3DCityDB. 
Moreover, as a demonstration of the capability of this methodology for integrating CityGML data with other datasets, we collect OSM data in the same test area. The selection of OSM data as an example is because it is one of the most popular crowdsourcing data worldwide and it contains complementary spatial and semantic information with CityGML data that can be combined for interesting queries and applications.
%
%
To show the usefulness of the generated KG, we collect real-world geospatial analytical tasks and formulate them as intuitive GeoSPARQL queries, which show a high degree of expressiveness.
%
Finally, we test three popular KG systems, i.e., Ontop, Apache Jena, and GraphDB, and confirm that the queries can be evaluated efficiently.

\section{Related work}
\label{sec:related_work}

\subsection{CityGML}

CityGML is a data model and exchange format for 3D digital modeling of cities and landscapes \cite{kolbe2009representing}.  The main advantage of CityGML in comparison to other data formats is that if offers the possibility to integrate semantic information within 3D city models. In 2008, CityGML became international standard in the Open Geospatial Consortium (OGC)\cite{citygml}.
Since then, CityGML draws more and more attention from mapping authorities, industries and academic societies. Nowadays, it is widely used for different applications in many countries and regions.

Aiming to modeling cities in 3D in the digital world, CityGML covers almost all types of features that could appear in urban area, namely, Building, Water body, Terrain, Transportation, Bridge, City Furniture, Land Use, Tunnel, etc. These features are organized into modules in CityGML. Although CityGML defines levels of detail (LoDs) for all types of features, the LoD of building objects is the mostly agreed and recognized concept in the 3D city modeling community.

\begin{figure}
  \centering
  \includegraphics[width=\textwidth]{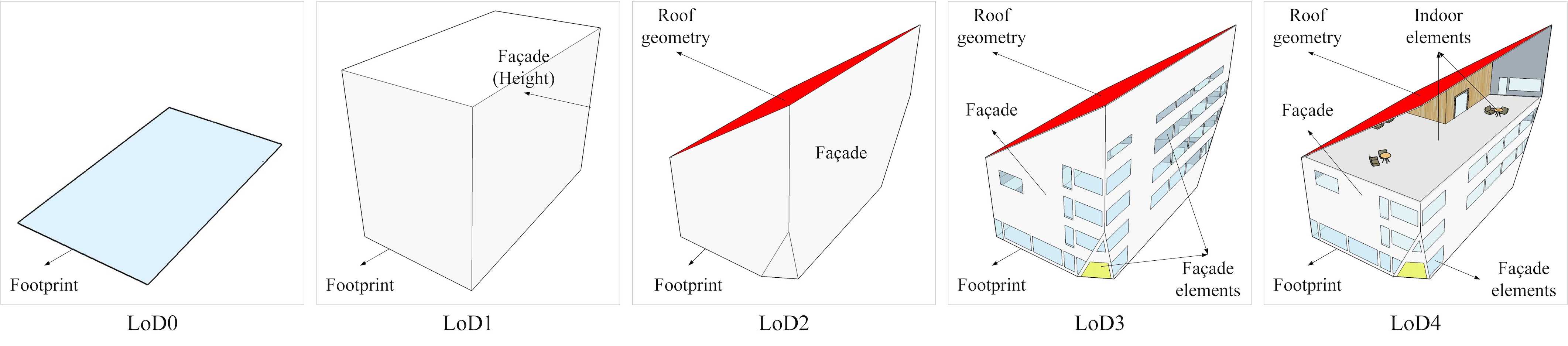}
  \caption{Five LoDs in CityGML}
  \label{fig:lods}
\end{figure}

In total, there are 5 LoDs defined for building objects in CityGML, ranging from coarse model (LoD0) to very detailed models (LoD4) with geometries and semantic information. 
As denoted in Figure~\ref{fig:lods} , 
an LoD0 building model in CityGML is actually a 2D footprint in a closed polygon which is semantically indicated as building object and can be added with various attributes. An LoD1 building in CityGML is a block model with height information, while an LoD2 building model needs have detailed roof structure but with walls as extruded 3D objects from 2D footprint. In LoD3 building models, architectural details on facades, such as windows, doors and other elements can be modeled in addition to LoD2 models. For a more further step, interior objects can be modeled in LoD4.

CityGML is very powerful to model 3D cities with rich semantic information. However, its complex and hierarchical structure, and also the interoperability issues create difficulties and complexity when transforming and decoding it for visualization and application scenarios~\cite{Noardo19b}. In order to overcome this issue, CityJSON was developed~\cite{ledoux2019cityjson} by combining the advantages of JSON and CityGML. In other words, CityJSON can be regarded as a JSON implementation of a subset of CityGML version 2.0. In 2021, CityJSON was accepted as OGC standard. Currently, people are working to adjust CityJSON to the CityGML 3.0 conceptual model.

\subsection{Knowledge Graphs and Geospatial Knowledge Graphs}

The Semantic Web research area is concerned with the challenges posed by data with a complex structure and associated knowledge. A prominent technology within the Semantic Web is that of knowledge graphs (KGs) \cite{Hogan_2021}, where data is structured in the form of a graph.
Domain objects and data values are represented as nodes of such a graph, and properties of objects are encoded
as edges. At the core of solutions based on KGs we typically have an ontology to provide semantics to the data.
In computer science, the term “ontology” denotes a concrete artifact that conceptualizes a domain of interest
and allows one to view the information and data relevant for that domain in a sharable and coherent way.
To simplify the sharing and reuse of
ontologies, the World Wide Web Consortium (W3C)\footnote{\url{https://www.w3.org/}}
has defined standard languages. We refer here to the Resource
Description Framework (RDF) \cite{Manola2004}, providing a simple mechanism to represent the data in a certain domain, and
Web Ontology Language (OWL)\cite{Hitzler-2009aa}, providing a very rich language to encode complex knowledge in the domain
of interest.

GeoSPARQL is a standard by OGC for representation and querying of Geospatial
KGs~\cite{OGC-GeoSPARQL}. %
GeoSPARQL provides a topological ontology in RDFS/OWL for
representation using Geography Markup Language (GML) and well-known
text representation of geometry (WKT) literals, and topological
relationship vocabularies and ontologies for qualitative reasoning.
GeoSPARQL also provides a SPARQL query interface using a set of topological SPARQL extension functions for quantitative reasoning.

Geospatial KGs are often converted from geospatial data sources which are  
stored in spatial databases or other popular used formats like Shapefiles. 
A systematic approach to such conversion is the \textit{ntology-based data access} (OBDA)
paradigm, which enables end users to access data sources through a domain
ontology. Typically, this ontology imports the GeoSPARQL ontology, and
is semantically linked to the data sources by means of a mapping, which is expressed in the R2RML language~\cite{r2rml} standardized by the W3C.
OBDA can be realized in a materialized or virtual fashion:
\begin{itemize}
\item In the \textit{Materialized Knowledge Graph
  (MKG)} approach, the original data sources are first materialized as RDF Graphs using systems, e.g., GeoTriples~\cite{KYZIRAKOS201816} and Ontop~\cite{2020-iswc-ontop},
  and then are loaded into RDF stores that support geospatial KGs, e.g.,
  Apache Jena\footnote{https://jena.apache.org/},
  GraphDB\footnote{https://graphdb.ontotext.com/}, and
  Stardog\footnote{https://www.stardog.com/}.
\item 
In the  \textit{Virtual Knowledge Graph (VKG)} approach, the content of KG is not generated but can be kept
virtual.  The ontology and mapping together, called a \textit{VKG
Specification}, exposes the underlying data source as a virtual RDF
graph, and makes it accessible at query time. For example,
Ontop~\cite{2020-iswc-ontop} is a popular VKG system that supports
GeoSPARQL.
\end{itemize}

One of the most famous Geospatial KG projects is
LinkedGeoData~\cite{SLHA11,2021-jws-lgd}, which mostly relies on the
VKG approach to expose the OSM data as Geospatial KGs.
To evaluate the systems for Geospatial KGs, Jovanovik \textit{et al.}
proposed a GeoSPARQL Compliance
Benchmark~\cite{DBLP:journals/ijgi/JovanovikHS21}, and
Li \textit{et al.}~\cite{DBLP:journals/urban/LiWWGT22} carried out an
extensive evaluation of several Geospatial RDF triple stores, showing
that both MKG and VKG systems have their advantages.
%





\subsection{Semantic technologies for 3D city models}

There have been early attempts to convert CityGML to knowledge graphs. In~\cite{CHADZYNSKI2021100106}, the authors use a straightforward ad-hoc implementation. They first refined an existing CityGML ontology from the University of Geneva, and then extended a corresponding data transformation tool that was originally designed to work alongside CityGML, which allowed for the transformation of original data into a form of semantic triples. Various scalable technologies for this semantic data storage were compared and Blazegraph was chosen due to the required geospatial search functionality. 
Many applications in the context of urban informatics require detailed information about the physical urban environment, which requires the integration of 3D city data with other data sources.
The work \cite{Ahmadian-osm-citygml-integration} studies the integration of OpenStreetMap and CityGML using the formal concept analysis approach.
Most work focuses on the integration of CityGML and BIM. The work~\cite{isprs-annals-VI-4-W1-2020-93-2020} proposed two approaches to integrate and reconcile city models and BIM in the context of solar energy simulations, where BIM data is stored in IFC and the city model in CityGML (LOD2). The first approach is to perform a schema matching in an ETL tool, so as to convert and import
window information from the IFC file into the CityGML model to create a LOD2-3 building model. In the second approach, they adopted a semantic web approach, in which both the BIM and city models are transformed into knowledge graphs
(linked data). City models and BIM utilize their respective but interlinked domain ontologies. Particularly, two ontologies are
investigated for BIM data, i.e., the ifcOWL ontology and the building topology ontology (BOT).
%

\section{Methodology}
\label{sec:methodology}

\begin{figure}[hbt!]
\vspace{-0.5em}
\centering
\includegraphics[width=0.9\textwidth]{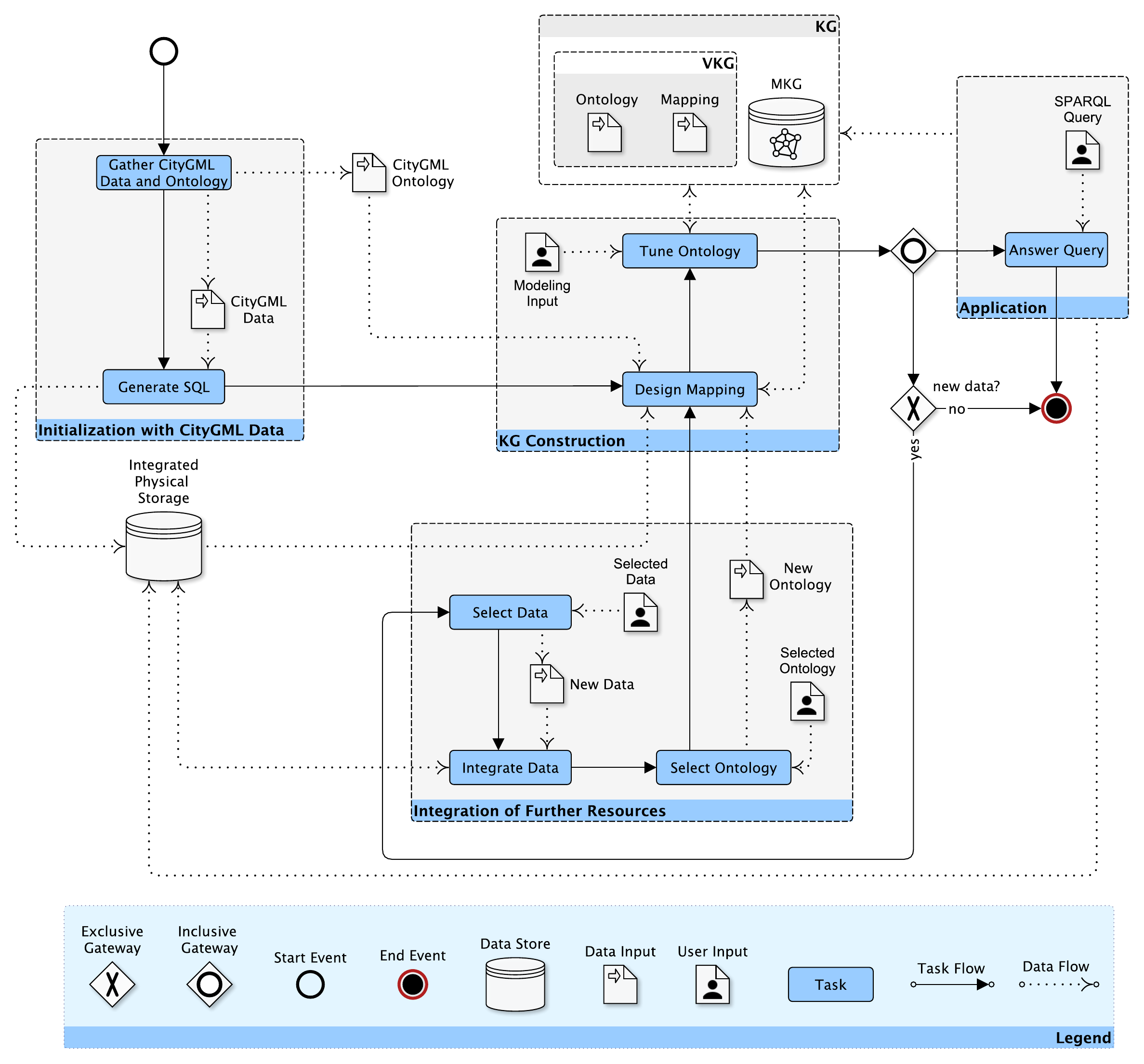}
\caption{CityGML KG Creation and Evolution: Overall view} \label{fig:pipeline}
\end{figure}

In what follows we describe the approach we adopted for generating the target KG from CityGML data and integrating other data sources. An overall view, denoted as a pipeline, is unveiled through the utilization of the \textit{Business Process Model and Notation (BPMN)}\footnote{Note that BPMN is a conceptual modeling language adopted to represent tasks and procedures within a system. To get more information about BPMN, the authors refer the readers to \cite{white2004introduction}.} diagram represented in \autoref{fig:pipeline}. 

The scope of the whole process is twofold. Firstly, it allows for the generation of a KG to support query answering over CityGML data. Secondly, it allows for the evolution of the created KG by importing new data and knowledge, thus enabling the possibility of extending question-answering services. 

Let us delve into the specific aspects of the approach. As shown by \autoref{fig:pipeline} we have four main groups of elements, which we may also call ``phases'', namely
\begin{inparaenum}[\itshape (i)]
\item \textit{Initialization with CityGML data} (hereafter ``Initialization''), 
\item \textit{KG Construction},
\item \textit{Integration of further resources} (hereafter ``Integration''), and
\item \textit{Application}.
\end{inparaenum}
 These phases are composed of sub-tasks or steps, which, in turn, may receive and/or produce different kinds of data. Differently, the \textit{KG} group represents the final output to be used as main support for the query-answering activities, which are represented in the \textit{Application} phase group. 
The output KG can be created either as a VKG or MKG. The VKG contains two sub-components, namely an \textit{ontology}, and a \textit{mapping} function which are used to generate the RDF triples from the physical storage on demand. Representing the KG as an MKG instead eliminates the need to maintain a virtualized pipeline for RDF data, with the trade-off of larger space requirements and the need to rematerialize the RDF triples every time the source data changes. All these components can be then evolved through the steps composing the \textit{Integration} phase. 

In this setting, the \textit{Initialization} phase has the primary goal of generating a reference data storage out of \textit{CityGML} data and, also, providing the reference CityGML ontology to be used as a baseline version of the knowledge graph. For the creation of the reference data storage \textit{the input CityGML dataset is embedded into a relational database} (See Generate SQL in \autoref{fig:pipeline}), mapping each row in the data into entities and columns in specific information fields so that the data can be then queried, retrieved, stored, and possibly updated. Note that, to address this step, albeit multiple automatic and comprehensive solutions are available, some \textit{ad hoc} customization sub-steps may be involved. This is due to the fact that the output physical storage of this phase must be compliant with the technology used to generate the knowledge graph in the following phase.
For instance, if the solution for generating SQL databases from the input data uses XML attributes containing (semi-)structured information,
a customization step would be needed to generate multiple fields out of each XML attribute\footnote{To have more information about implementation issues please refer to Section~\ref{sec:architecture}}. 

The second phase  \textit{KG Construction} crafts a KG that can be used as a reference point for the query-answering activities. Note that this step can be iterated multiple times.
The first time is after the initialization phase, and the inputs for KG construction will naturally be the \textit{CityGML} ontology, the \textit{Physical Storage} hosting the \textit{CityGML} data, and the \textit{mapping} that is necessary to connect the former to the latter. In the following times, the inputs of this phase are the result of the integration phase.
Either way, the KG Construction phase is mainly concerned with the definition of the ontology and the related mappings, 
with the main scope of
\begin{inparaenum}[\itshape (i)]
\item defining the set of concepts, relationships, and properties
  within the reference domain of knowledge;
\item capturing the meaning and semantics of the stored information, by enabling extended reasoning and inference capabilities; and
\item fostering interoperability among the data sources to be integrated.
\end{inparaenum}
A key aspect here is also to find an already existing ontology that covers as much of the semantics of the selected data as possible. Once the ontology is selected a mapping step is performed. The database generated through the initialization phase is then aligned with the ontology concepts.
When the information cannot be straightforwardly mapped, a manual intervention is required. This mainly involves a modification of the selected ontology in order to properly account for the information in the physical storage. For example, if a property that is present in the input dataset is not present in the ontology, the ontology can be manually extended with the required property. Once the ontology is tuned according to the dataset requirements, the KG is created and ready to support the query-answering application. 

After the adoption of both the creation and KG construction phases, as we anticipated above, we already have a reference KG enabling query answering. However, at this point, our approach allows for the integration of more data sources and then for the creation of an extended KG, on top of the previous version. This evolution is addressed through the \textit{Integration} phase. Here the main tasks are
\begin{inparaenum}[\itshape (i)]
\item the selection of new data by the user,
\item the integration of the new data with the existing physical storage, and 
\item the selection of a new ontology or new ontological information by the user, in order to account for the newly integrated data. 
\end{inparaenum}
Task \textit{(ii)} takes place with the support of an automatic component that involves two sub-steps, namely \textit{(ii.a)} \textit{post-processing} where heterogeneous geo-object types, data formats, and coordinate reference systems (CRS) are harmonized and unified and \textit{(ii.b)} \textit{spatial matching} where the similarity between geo-entities from different datasets is calculated.

Finally, the phase named \textit{Application} is dedicated to using the output KG. Here the user, via an \textit{ad hoc} interface, is enabled to request information via SPARQL queries, which can potentially be returned in either textual or visual format. 


\section{System Architecture}
\label{sec:architecture}
\begin{figure}[hbt!]
\vspace*{-0.5em}
\centering
\includegraphics[width=1\textwidth]{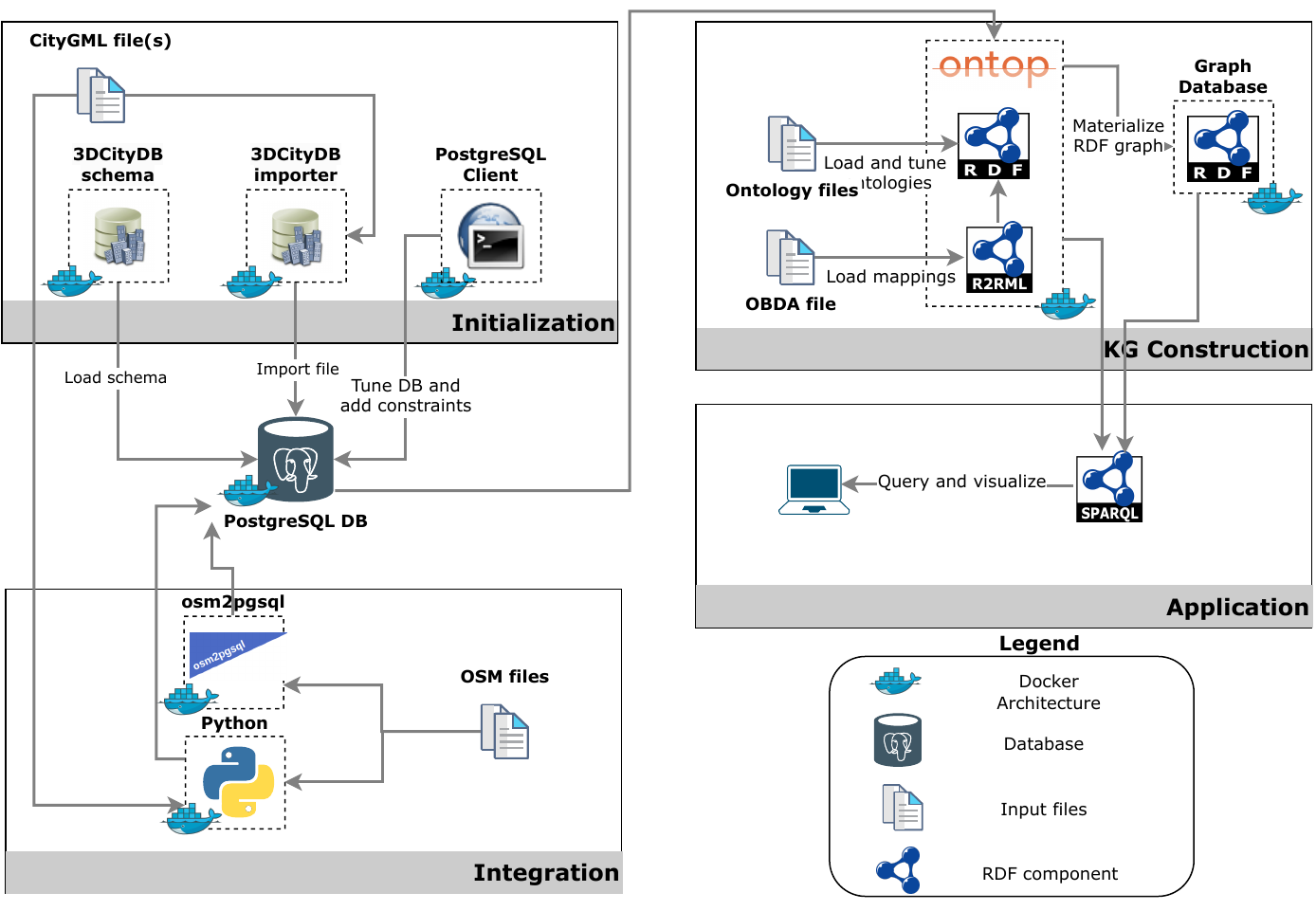}
\caption{VKG over CityGML: Architecture} \label{fig:architecture}
\end{figure}

%
In this section, we describe how the conceptual methodology framework in Section~\ref{sec:methodology} can be realized in a concrete system. As shown in \autoref{fig:architecture},
in the context of our particular application scenario, we examine two typical 3D and 2D geospatial data sources, namely \textit{CityGML} and \textit{OpenStreetMap (OSM)}. CityGML data is used throughout the whole phases, while OSM data, one of the most comprehensive and widely used geospatial data sources, is adopted to illustrate 
the \textit{Integration} phase. 
More specifically, we use LOD2 CityGML building data retrieved from the Bavarian Open Data portal\footnote{\url{https://geodaten.bayern.de/opengeodata/OpenDataDetail.html?pn=lod2}} in the central area of Munich, Germany and OSM data in the same area as a demonstration. Please refer to Section~\ref{sec:experimental-set-up} for further details on the test area and datasets. 

\subsection{Initialization with CityGML data}

The initialization phase generates a reference data storage out of CityGML data. 
We selected the \textit{3DCityDB schema} as the preferred solution to import CityGML data into an SQL database. 
3DCityDB as a software solution provides both a predefined SQL schema\footnote{\url{https://github.com/3dcitydb/3dcitydb}} and importer-exporter\footnote{\url{https://github.com/3dcitydb/importer-exporter}} tool which can process the cumulative addition of an arbitrary number of CityGML files into the database. 
The software provides the option to use PostgreSQL or Oracle as the backend for the relational data storage. In this work, we chose PostgreSQL in particular because it is open source and its geospatial extension PostGIS is renowned for its high adoption and maturity in the geospatial domain. 

An excerpt of the table \texttt{building} with three records is presented in \autoref{tab:citygmlRDBMS} (omitting any attributes with missing data). Every building is uniquely identified by the attribute \texttt{id} and has a corresponding LOD2 solid identifier \texttt{lod2\_solid\_id}. The latter is mapped onto its respective polyhedral surface serialization in the \texttt{surface\_geometry} table. Further sample data is provided in \autoref{fig:mapping_examples}.
\begin{table}[tbp]
    \centering
    \resizebox{\columnwidth}{!}{
\renewcommand{\arraystretch}{1.3}
    \begin{tabular}{llllll}
      \toprule
      \texttt{id}&\texttt{objectclass\_id}&\texttt{building\_root\_id}&\texttt{roof\_type}&\texttt{measured\_height}&\texttt{lod2\_solid\_id} \\\midrule
      10&26&10&1000&13.363&117\\ 
      54&26&54&3100&13.99&315\\ 
      248&26&248&1000&17.362&1258\\ \bottomrule
    \end{tabular}}
\caption{Excerpt from the CityGML \texttt{building} table}
  \label{tab:citygmlRDBMS}
\end{table}

Finally, we note that despite the comprehensive default SQL schema provided by 3DCityDB ,
a further step is needed to \textit{tune DB and add constraints}. For example, the attribute of \texttt{address} in the default database schema is encoded as XML strings and has to be decomposed into more specific attributes, e.g. administrative area, thoroughfare, etc. Relevant constraints like primary and foreign keys are added to enhance the efficiency.

\subsection{Knowledge Graph Construction}

In this architecture, we support constructing KGs in both VKG and MKG fashions. We first construct the VKG utilizing the \emph{Ontop} system. 
%
The main activity is to develop/refine the \textit{ontology} and create \textit{mappings} to link the terms (classes and properties) in the ontology to the data sources.
We will also use Ontop to materialize the RDF triples into an MKG, and load it to a triple store system like Jena or GraphDB.

\begin{figure}[hbt!]
    \centering
    \includegraphics[width=1\textwidth]{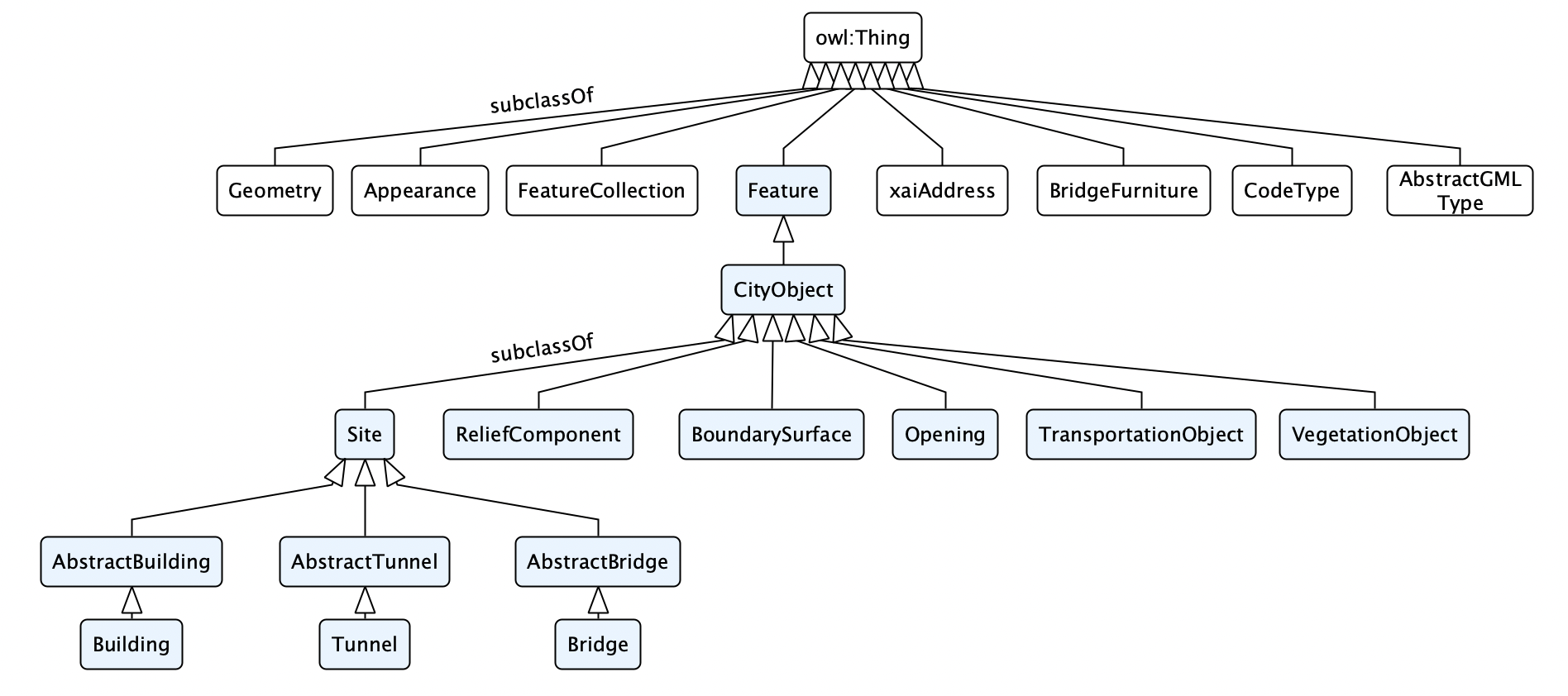}
    \caption{A subset of the concepts in the CityGML ontology with the subtree of the class \texttt{Feature} highlighted}%
    \label{fig:protege_screenshot}%
\end{figure}

\smallskip
\noindent\textit{Ontology.}
We adopt the most prominent and well-known \textit{CityGML} ontology version~2.0\footnote{\url{https://cui.unige.ch/isi/ke/ontologies}} developed by the University of Geneva 
for the ontology component of the KG construction phase. 
We first \textit{load and tune ontologies} from the ontology provided. Following validation of the ontology, there were 92 declarations of object and data properties using the same IRIs, which makes the ontology invalid.
The same inconsistencies have been diagnosed in previous research \cite{CHADZYNSKI2021100106}. We tackled the issue in a similar fashion resolving any inconsistencies manually depending on the most intuitive definition of each property. In order to review and resolve these inconsistencies and get an overview of the collection of ontologies we utilized the open-source tool \protege version 5.6.1. \autoref{fig:protege_screenshot} shows part of the top-level classes of the selected ontology and the sub-tree starting from \texttt{Feature} and \texttt{Building}, which are of primary interest in this study.

We also note that in addition to the primary CityGML ontology, auxiliary ontologies were also specifically \texttt{geosparql}, \texttt{gml}, \texttt{sf}, \texttt{sosa}, \texttt{core}, \texttt{dublin\_core\_elements}. These ontologies do not define additional concepts but rather help define constraints within the CityGML ontology. The GeoSPARQL ontology~\cite{OGC-GeoSPARQL}, for instance, allows the differentiation of geometric classes such as polyhedral surfaces from standard surfaces while complying with the standard OGC recommendations.


\smallskip\noindent\textit{Mappings.}
Mapping design is the most crucial user-centric step in generating a VKG. Individual RDB2RDF mappings exploit attributes from the PostGIS database to populate the RDF graph of CityGML. Consequently, SQL queries have to be written to individually map 3DCityDB attributes to their respective ontological concepts. Due to the limitation of LOD2 Bavarian data (and any open CityGML data we can find), which contains exclusively buildings, and lack of any complementary real-world LOD3 files, many 3DCityDB tables are empty. Therefore, while for completeness purposes any column from the 3DCityDB schema that could be mapped has been mapped to an ontological concept, in practice no triples can be generated from many of these mappings. 

\begin{figure}[hbt!]
\vspace{-0.5em}
\centering

\subfloat[Mapping \texttt{Building}]{
  \includegraphics[width=0.9\textwidth]{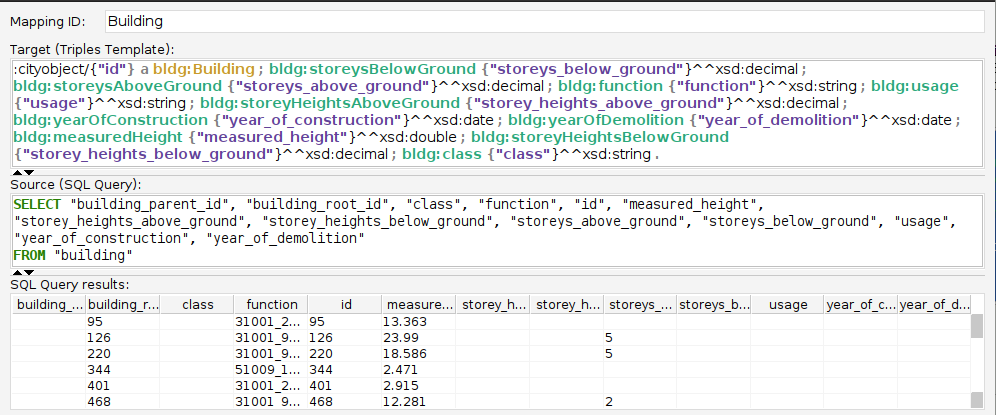}
  \label{fig:ontopplugin_screenshot_building}
} 

\subfloat[Mapping \texttt{Building - LoD2Solid}]{
  \includegraphics[width=0.9\textwidth]{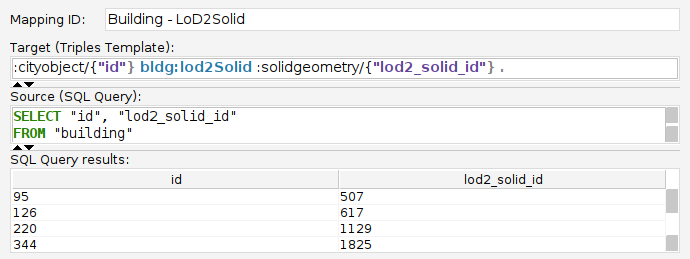}
  \label{fig:ontopplugin_screenshot_buildingtosolid}
} 

\subfloat[Mapping \texttt{Solid Geometry}]{
  \includegraphics[width=0.9\textwidth]{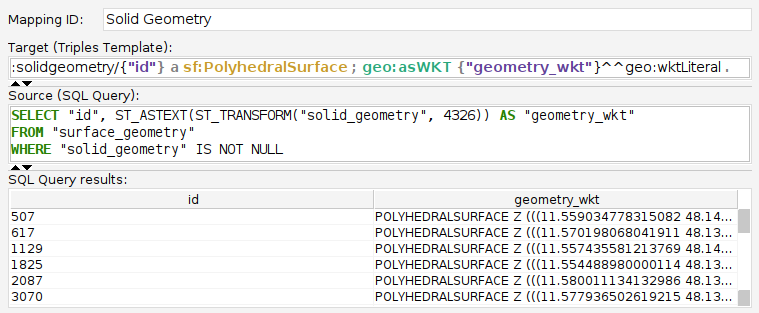}
  \label{fig:ontopplugin_screenshot_solidgeometry}} 
    
\caption{Three example mappings in Ontop \protege Plugin}
\label{fig:mapping_examples}

\end{figure}

A mapping consists of three components: a mapping ID, a source, and a target.  A mapping ID is an arbitrary but unique mapping identifier. A source refers to an SQL query expressed over a relational database for retrieving data. A target is RDF triple pattern(s) that uses the answer variables from the preceding SQL query as placeholders. Three example mappings written in Ontop \protege plugin editor are illustrated in \autoref{fig:mapping_examples} to respectively define a building, link a building to its respective solid geometry, and define the serialization of that geometry.
More specifically, in the first mapping in Figure~\ref{fig:ontopplugin_screenshot_building}, the class \texttt{Building} is mapped with its respective data properties such as building height, storeys above and below ground, function, year of construction, etc.
The second mapping in~\autoref{fig:ontopplugin_screenshot_buildingtosolid} shows that object property \texttt{bldg:lod2solid} links any building with its respective solid geometry identifier. We distinguish between a solid geometry class and its respective serializations which are properties of the class. 
In the third mapping in~\autoref{fig:ontopplugin_screenshot_solidgeometry}, class \texttt{sf:PolyhedralSurface} defines objects of type polyhedral surface and their respective Well-Known Text (WKT) geometry serialization.

\begin{figure}
  \vspace{-0.5em} \centering

  \subfloat[Triples for CityGML]{
    \includegraphics[width=0.8\linewidth]{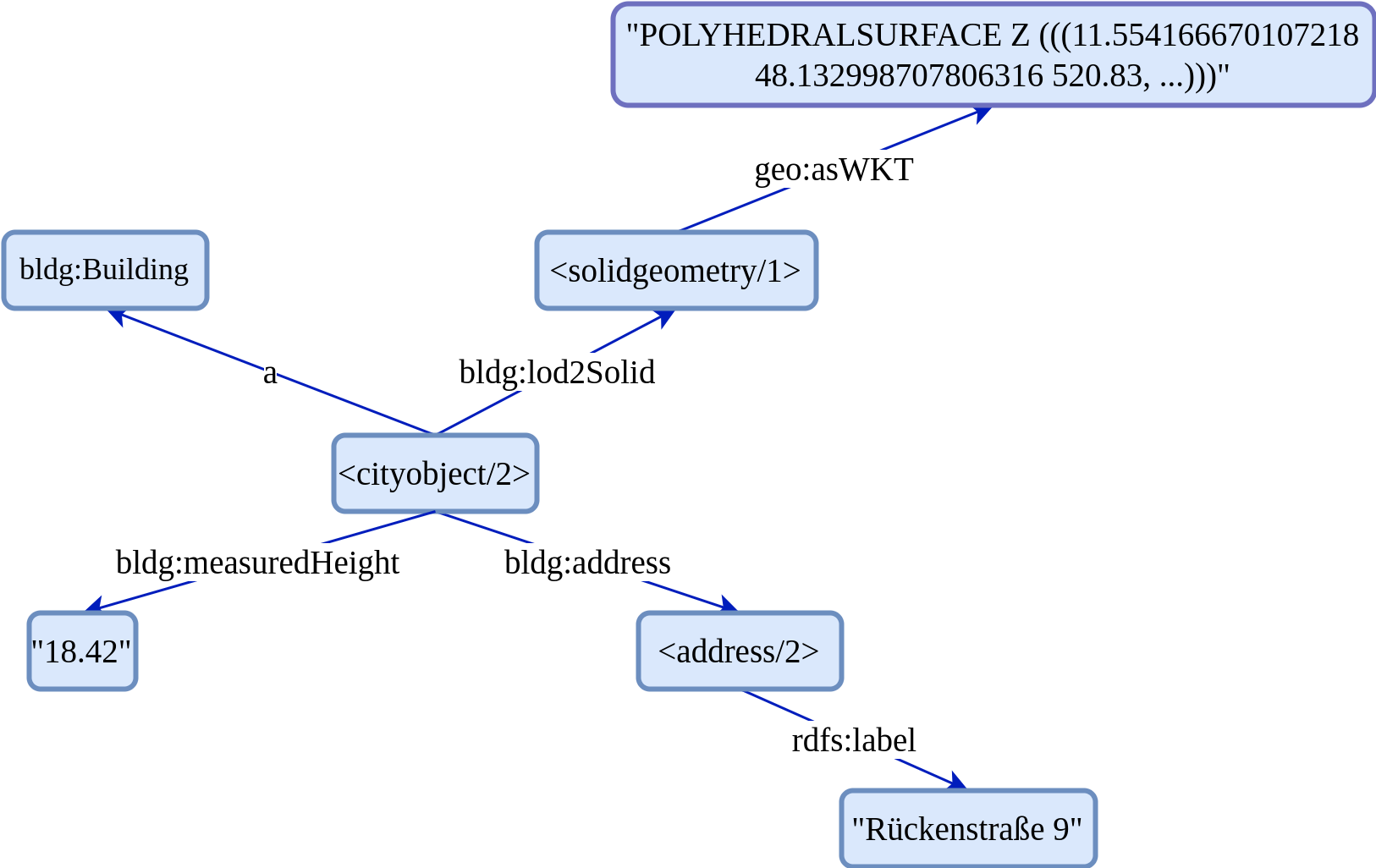}\label{fig:example-triples-citygml}}

  \subfloat[Triples for CityGML and OSM]{
    \includegraphics[width=0.8\linewidth]{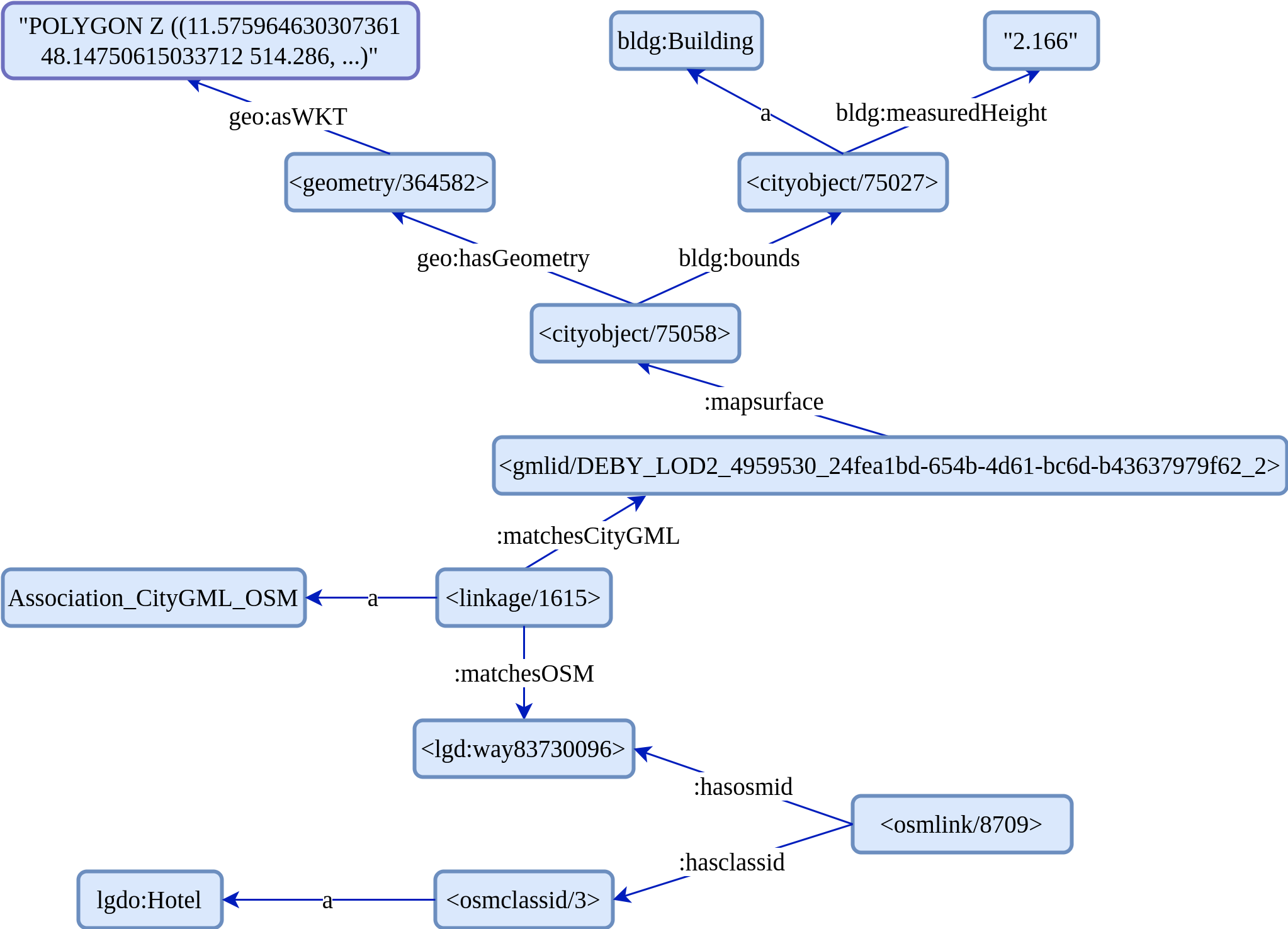}\label{fig:example-triples-osm-citygml}}

  \caption{Example of triples in the Knowledge Graphs}
  \label{fig:example-triples-citygml-osm}
\end{figure}

\smallskip
\noindent\textit{KG Materialization}
With the completion of all the ontology and mappings, the CityGML VKG has been successfully created. This VKG can be queried already by Ontop, or be materialized to use native MKG systems.
The MKG is constructed by utilizing the functionality of Ontop to generate RDF triples or data assertions based on the ontology, mappings, and physical storage, which were discussed in detail in the previous section.
The resulting triples can be in turn loaded into RDF triple stores like GraphDB, Apache Jena, RDF4J, and similar tools to facilitate query answering via SPARQL. Further details on how these triple stores can exploit the materialized MKG are provided in \autoref{sec:experiments}. Examples of sets of triples generated by our application pipeline are depicted in \autoref{fig:example-triples-citygml-osm}. They reflect the triples generated from the three mappings described in the previous section. Specifically, \autoref{fig:example-triples-citygml} shows a sub-graph from the example mappings representing a building and its LOD2 geometry and address. 

\subsection{Integration of OSM data} 


Below we describe the steps of integrating further geospatial data sources in our architecture by using OSM data as an example.
For any other generic geospatial data, the integration task will be contingent on the type of data we wish to integrate and any existing popular ontologies. For example, for what concerns \textit{OSM data}, as mentioned in Section \ref{sec:related_work}, the LinkedGeoData project already leveraged the loading of OSM data into PostgreSQL and developed ontology and mapping, which can be reused in this work. What's missing is the linking between CityGML and OSM data at both the data level and the ontology level. This requires computing the correspondence between the data items, i.e. buildings, between these two data sets, and creating additional suitable mappings and ontological axioms to capture these correspondences.
As anticipated above, in order to handle this heterogeneity issue we leverage an entity resolution step that produces a reference linkage table to be used for the generation of the output physical storage as PostgreSQL DB.
Below we present a geometry-based method for linking entities in CityGML and OSM data sources and how to incorporate the results in the KG. 

\subsubsection{OSM and CityGML data linking}



Because of the heterogeneity between the CityGML and OSM datasets, we cannot expect that the resulting data linking is always 1:1. 
The building information of OSM data mainly consists of the building footprint layer (polygons) and the point of interest (POI) layer (points).
In contrast, the CityGML data is 3D, hence we primarily rely on the ground surfaces of the CityGML buildings.
The CityGML dataset normally has more detailed information about the buildings.
In particular, CityGML buildings frequently encompass minor ancillary features like stairs and garages, which are often absent in OSM building footprints, and may lead to a n:1 matching result.

We propose a three-step method of data linking:
\begin{enumerate}[\itshape (1)]
\item Computing direct spatial correspondence using CityGML ground surfaces and OSM polygons,
\item Exploiting the adjacent ground surfaces in CityGML to enrich the results, and
\item Matching OSM POI points with CityGML buildings.
\end{enumerate}

Note that more sophisticated approaches, \textit{e.g.}, formal concept analysis~\cite{Ahmadian-osm-citygml-integration}, can be adopted in this architecture as well, but to simplify the presentation, we only use the geometry-based approach in this work. 





\begin{figure}[htb]
\centering
\includegraphics[width=\textwidth]{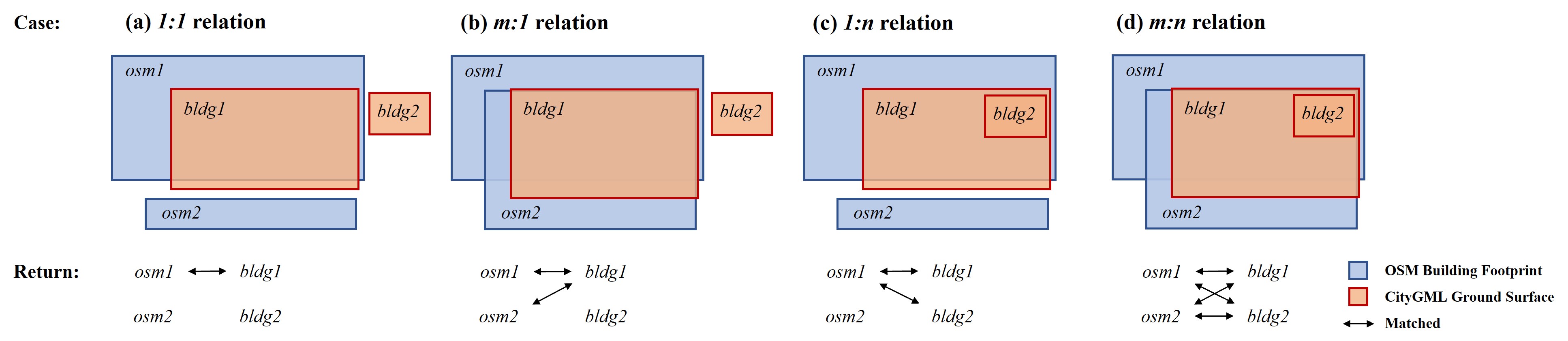}
\caption{Schematic diagram of the four spatial matching relations, namely (a) 1:1 relation, (b) m:1 relation, (c) 1:n relation, and (d) m:n relation.}
\label{fig:spatial_match}
\end{figure}

\smallskip
\noindent\textit{(1) Spatial matching.}
\label{sec:step1}
Since the CityGML data in this study only contains building information, we refer to the linking of CityGML and OSM polygons as the linking of building information between them.  Given any CityGML building ($bldg$) and OSM polygon ($osm$), their direct spatial correspondences are identified based on \autoref{match_equation}~\cite{fan2014quality,liu2023building}, derived from individual areas of any CityGML and OSM polygons ($\mathrm{Area}\left({bldg}_i\right)$, $\mathrm{Area}\left({osm}_j\right)$).
    \begin{equation}
        \frac{\mathrm{Area}\left({osm}_i\cap {bldg}_j\right)}{\min \left(\mathrm{Area}\left({osm}_i\right),\mathrm{Area}\left({bldg}_j\right)\right)}\ge t
    \label{match_equation}
    \end{equation}
where $\mathrm{Area}\left({osm}_i\cap {bldg}_j\right)$ represents the overlapping area of the \textit{i}-th OSM polygon and the \textit{j}-th CityGML polygon;
$t$ is an empirical hyperparameter, which can be adjusted based on the spatial consistency between the two datasets.
%
Following~\cite{fan2014quality}, there are four possible matching results based on the ratio: 
 1:1, 1:n, m:1, m:n (examples illustrated in Figure~\ref{fig:spatial_match}).
Relation 1:1 indicates that an OSM building and a CityGML building are uniquely matched with each other (Figure~\ref{fig:spatial_match}(a)).
Relation m:1 represents multiple OSM buildings matching with one CityGML building (Figure~\ref{fig:spatial_match}(b) and relation 1:n  the opposite case (Figure \ref{fig:spatial_match}(c)).
Relation m:n represents at least two OSM buildings matched together with at least two CityGML buildings (Figure~\ref{fig:spatial_match}(d)).

\begin{figure}[htb]
\centering
\includegraphics[width=.65\textwidth]{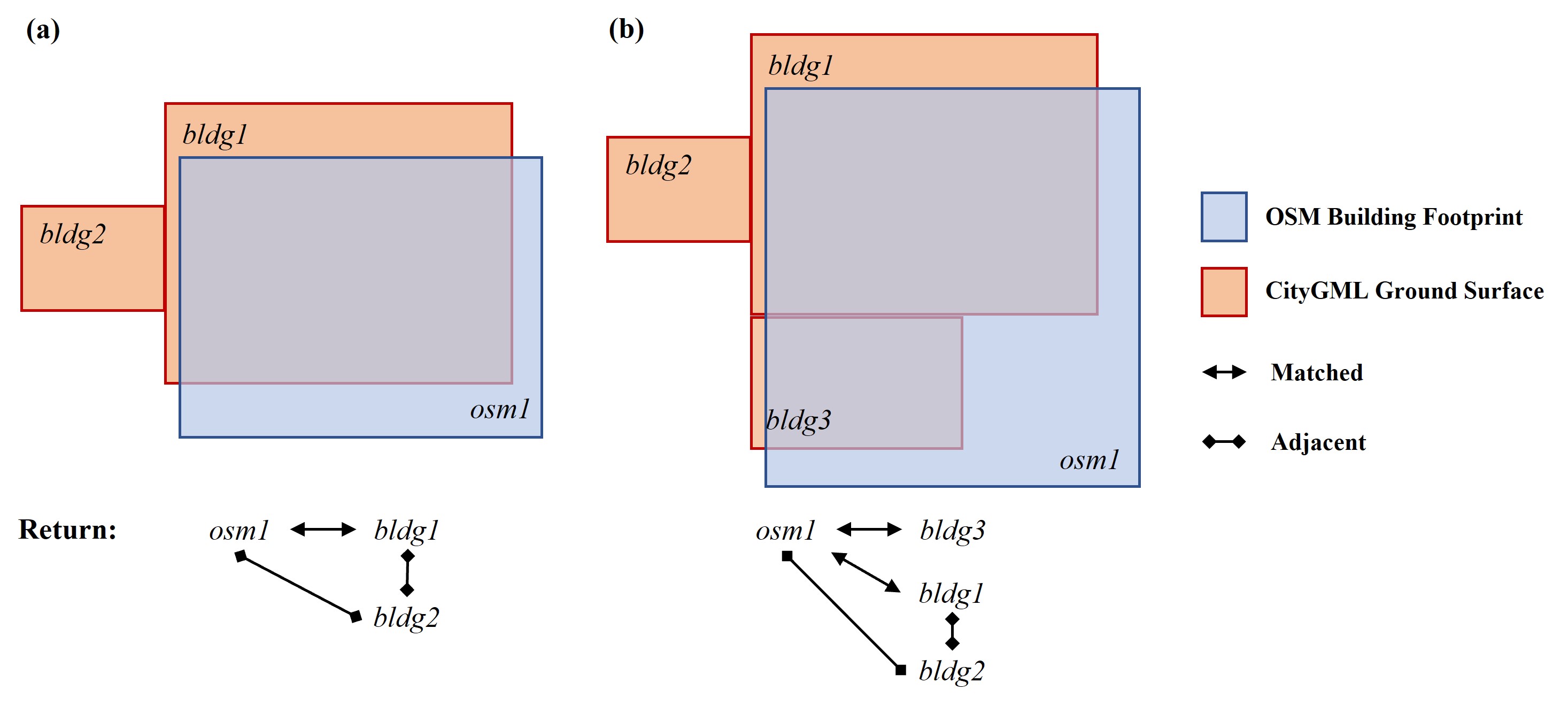}
\caption{Schematic diagram of the two adjacent identification situations in the OSM building \textit{osm1} perspective: (a) 1:1 relation with adjacency, and (b) 1:n relation with adjacency.} \label{spatial_adjacent}
\end{figure}

\smallskip
\noindent\textit{(2) Identification of adjacent polygons.}
This step specifically endeavors to include adjacent amenity objects as secondary matched (adjacent) relations, guaranteeing the inclusion of all amenities. In the examples illustrated in Figure~\ref{spatial_adjacent}, $bldg2$ and $osm1$ are matched as adjacent if the following three conditions are met: (a) $bldg1$ directly matches $osm1$, (b) $bldg1$ and $bldg2$ are adjacent, and (c) there is no other match for $bldg2$. 


\begin{figure}[h!]
\centering
\includegraphics[width=.65\textwidth]{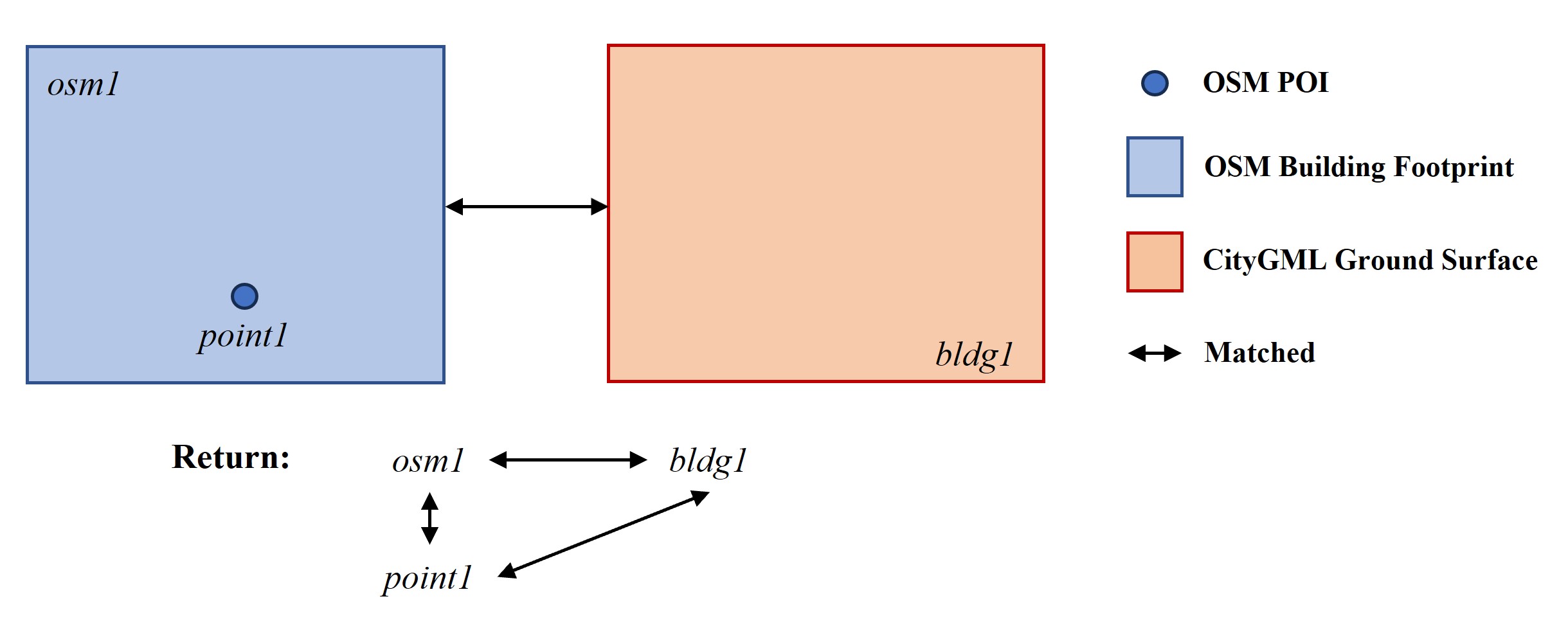}
\caption{Schematic diagram of the spatial match between OSM POIs and CityGML ground surfaces.}
\label{fig:OSM_POI_matching}
\end{figure}

\smallskip
\noindent\textit{(3) Matching OSM POIs with CityGML buildings.}
 This step aims to match the CityGML data with OSM POIs to enhance the semantic information of CityGML. The OSM POIs contain the place information in the buildings, e.g., various shops on different floors. The spatial locations of building-related POIs are based on the building footprints. Thus, we applied OSM's building footprints as a mediator to determine the spatial relationship between OSM POIs and CityGML buildings. As shown in Figure~\ref{fig:OSM_POI_matching}, given any POI, if the OSM building footprint where it is located matches a CityGML ground surface, the POI also matches the corresponding CityGML ground surface.


\smallskip

\paragraph{Empirical matching results}
As for the selected study area in Munich, the input OSM data contain 3,839 building footprints while the input CityGML data contain 5,728 ground surfaces. 
Previous studies have considered a minimum threshold $t$ of 30\% is necessary to determine the matching relationship \cite{fan2014quality,liu2023building}.
Empirically, we tried several settings and chose to set the tolerance threshold $t$ to 50\% in this case study according to the performance on both OSM and CityGML data.
In order to evaluate the correctness of spatial linking workflow, 50 randomly selected building polygons were manually examined, where all of them were correctly identified as matched or adjacent. 

\begin{figure}[h!]
\vspace{-0.5em}
\centering
\includegraphics[width=\textwidth]{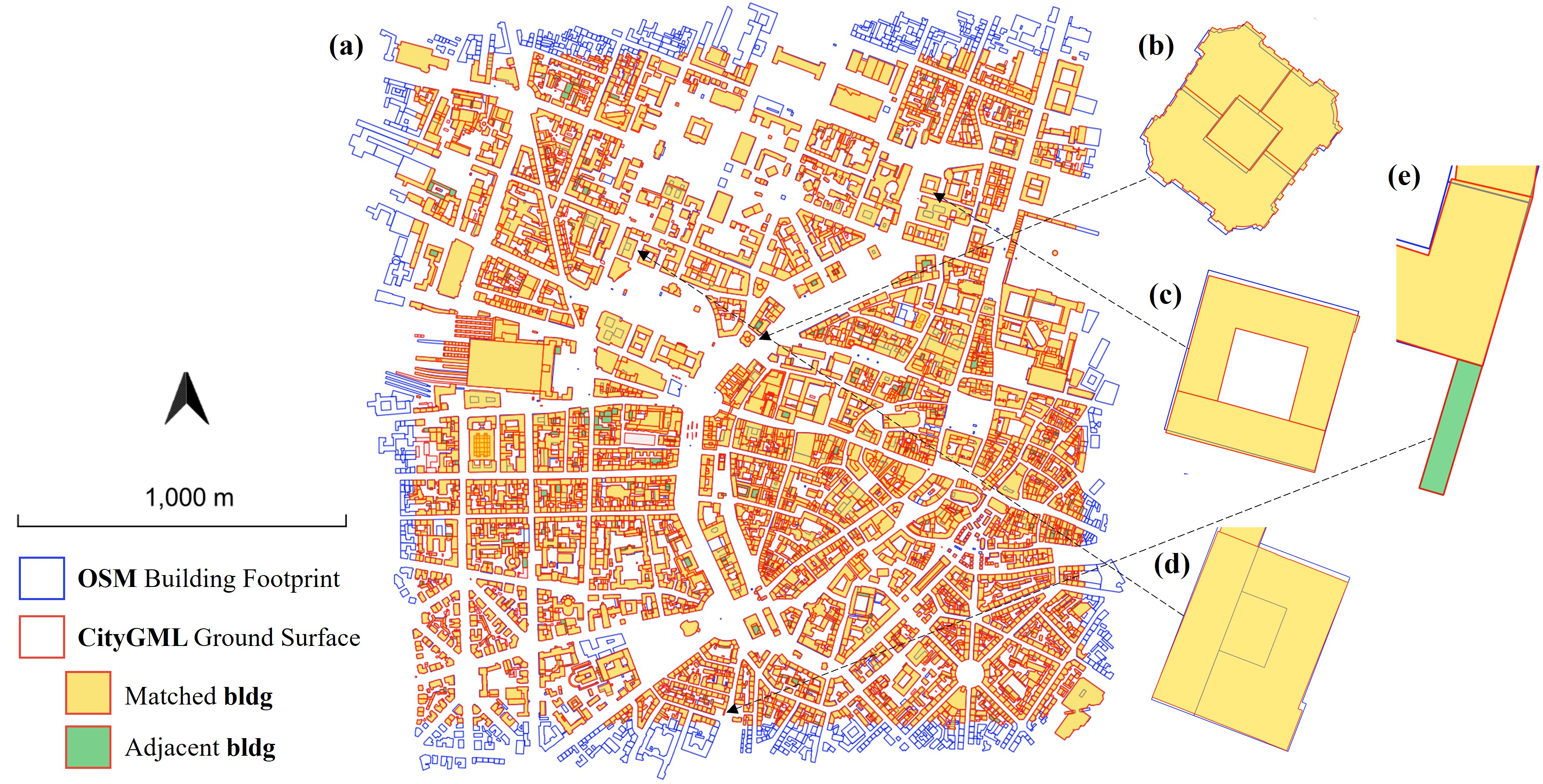}
\caption{Distribution of (a) the spatial integration result in the case study  and the cases of four common spatial relations, i.e., (b) 1:1 , (c) 1:n, (d) m:1, and (e) adjacent relations.} \label{fig:result_spatialMatch}
\end{figure}

\begin{figure}[h!]
\centering
\includegraphics[width=.9\textwidth]{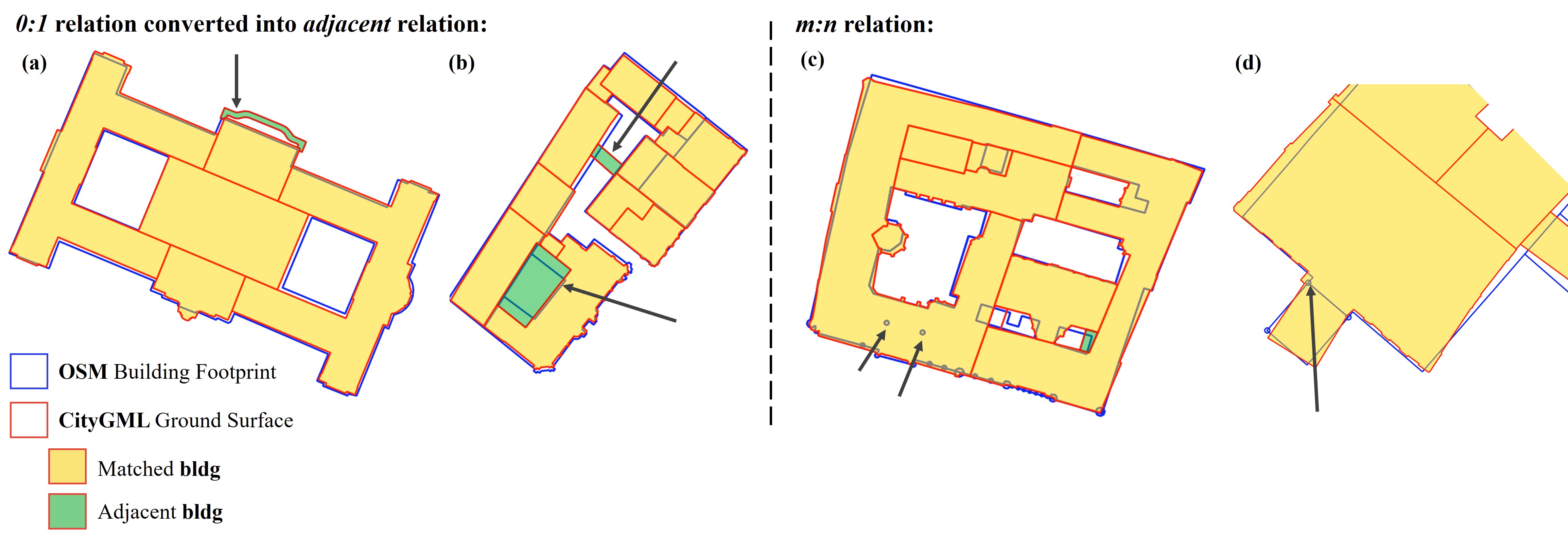}
\caption{Cases of 0:1 relation converted into adjacent relation (a-b) and cases of m:n relation (c-d).}
\label{fig:result_special_cases}
\end{figure}

Figure~\ref{fig:result_spatialMatch}(a) shows the results of \textit{Step 1 - spatial matching} between CityGML and OSM polygons in the study area. The majority of CityGML ground surfaces are successfully matched with OSM polygons. 
The 1:1 relations account for  42.92\% (2090 buildings). 
The 1:n relations (shown in Figure~\ref{fig:result_spatialMatch}(c)(d)) make up 16.20\% (789 buildings) while the m:1 relations only account for 5.87\% (286 buildings) of the total in \textit{Step 1}. This indicates that CityGML provides more detailed information, representing individual building accessories (e.g., staircases) as separate ground surfaces. Zero-to-one relations account for 21.17\% (1031 buildings) in \textit{Step 1} due to the same reason.
For instance, in \autoref{fig:result_special_cases}(a) and (b), the CityGML polygons pointed by the arrows should be included in their main buildings and linked with the corresponding OSM polygons. 

After \textit{Step 2}, these unmatched CityGML polygons are defined as matched by their adjacent OSM polygons. As a result, the 0:1 relations decrease from 21.17\% (1031 cases) to 5.54\% (270 cases), which demonstrates the necessity of including the adjacent structures as \textit{Step 2}.
Additionally, there is a 12.83\% (625 cases) occurrence of one-to-zero relations, where certain OSM buildings are absent in the CityGML data. This is mainly due to the slightly larger coverage of the downloaded OSM data compared to the CityGML data (as demonstrated in Figure~\ref{fig:result_spatialMatch}), ensuring that no CityGML buildings on the tile edges are missing in OSM. 
Ideally, overlapping polygons within a dataset's ground layer should be avoided. For instance, locations marked by arrows in Figure~\ref{fig:result_special_cases} (c) and (d) shouldn't serve as two buildings' foundations. The integration step is intentionally designed to account for such specialized relations. 
In this study area, only 49 cases of m:n relation (1\%) were identified.

As for \textit{step 3}, within the specified study area, we have successfully matched 2,718 OSM POIs with CityGML buildings.





\subsubsection{Modeling the linking results into the KG}
Integration at the relational database level needs to be further upstreamed to the ontology and knowledge graph level. Firstly, in order to query OSM concepts, a supplementary ontology, namely the \textit{LinkedGeoData (LGD)} ontology, originally defined by~\cite{SLHA11}, was adopted. LGD defines over 1,200 classes and 700 properties by leveraging the most ubiquitous tags present in OSM data. 
LGD enriches the CityGML KG with classes that represent OSM points of interest like hotels, residential buildings, and primary highways, as well as properties such as business opening hours and websites.



Secondly, there is a task to model at the KG level the connections between the still disjoint CityGML and OSM sub-KGs.
Utilizing existing owl terms like \texttt{owl:sameAs} is insufficient since we do not model identity or matches between individual buildings but other properties like adjacency, and potential associations between buildings could be enriched beyond that.

\begin{figure}
    \vspace{-0.5em}
    \centering
    \includegraphics[width=.95\textwidth]{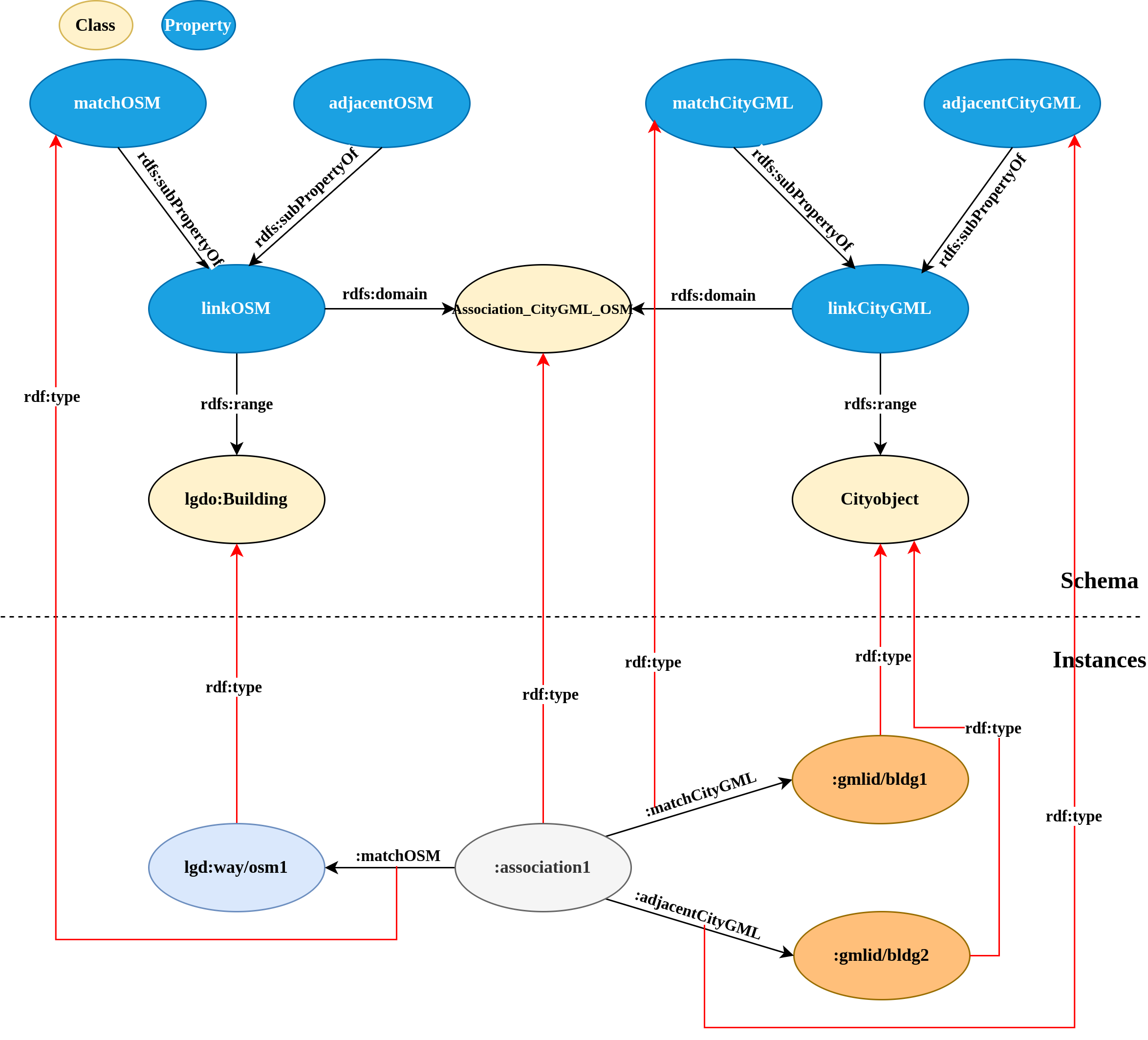}
    \caption{Modelling CityGML and OSM Association in the KG}
    \label{fig:association_citygml_schema_and_instances}
\end{figure}

Modeling the spatial matching results from the database to the KG necessitates the reification of the association between a CityGML building and OSM building. In practice this requires creating an additional class to represent this relationship which we define as \texttt{Association\_CityGML\_OSM}. This relationship allows the addition of further properties for \texttt{Association\_CityGML\_OSM}, e.g. it can now model both matched buildings and adjacent buildings.  An example of how an association would be expressed in the KG is shown in \autoref{fig:association_citygml_schema_and_instances}.
The upper part illustrates the schema (i.e., classes, properties, and their relations), and the lower part concrete instances corresponding to the example from \autoref{spatial_adjacent} (a) where e.g., an OSM building \texttt{lgdo:way/osm1} is linked to both a matching and adjacent CityGML building surface as defined respectively by \texttt{gmlid/bldg1} and \texttt{gmlid/bldg2}. Both the match and adjacency are modeled as subproperties of the linkage.




Note that in the example above, we use the instances from \autoref{spatial_adjacent} (a) where e.g. an OSM building \texttt{lgd:way/osm1} is linked to both a matching and adjacent CityGML building surface as defined respectively by \texttt{gmlid/bldg1} and \texttt{gmlid/bldg2}. Both the match and adjacency are modelled as subproperties of the linkage.

\section{Experiments}\label{sec:experiments}

In this section, we conduct a series of experiments in order to evaluate the following aspects:
\begin{enumerate}
    \item the expressiveness capabilities of the KG. We determine whether a KG constructed over CityGML data and further integrated with additional ad hoc data sources suffices in answering legitimate semantic queries designed by domain experts;
    \item the performance of the query evaluation with representative KG systems, including both VKG and MKG systems. 
      This will serve to determine whether results can be retrieved within a reasonable time frame from  domain experts.
\end{enumerate}

\noindent
The experiments are reproducible following the execution of the respective queries and setup described in the online appendix\footnote{\url{https://github.com/yuzzfeng/D2G2/tree/main/WP2\%20Thematic\%20Enrichment/linkage_demo}} as well as \autoref{sec:appendixA}.

\subsection{Experimental setup}
\label{sec:experimental-set-up}

The experiments are conducted on a normal laptop machine running 4 cores (Intel(R) Xeon(R) Gold 6154 CPU @ 3.00GHz), 16GB RAM, and 350GB SSD hard disk, running Ubuntu operating system.
The whole experiment environment has been setup as Docker containers to encapsulate all necessary software, which means experiments can be conducted under any operating system. 
For storing and querying CityGML data, we use the Docker image of 3DCityDB v4.1.0 that comes with PostgreSQL v15 and PostGIS v3.3,  corresponding to the latest versions available at the time.

Three KG systems that support  GeoSPARQL have been selected for the
experiments: one VKG system Ontop and two MKG systems (a.k.a triple
stores) Apache Jena and GraphDB. For GraphDB and Jena, a preparatory step of materializing all the triples is necessary. We carry this out using Ontop, and then load the file into Apache Jena and GraphDB as an input. Further descriptions of the evaluated systems are provided below:

\begin{itemize}[-]

\item Ontop\footnote{\url{https://ontop-vkg.org/}} is an open-source software project that focuses on providing a platform for efficient querying of relational databases using Semantic Web technologies, specifically the RDF data model, SPARQL query language, OWL 2 QL ontology, and R2RML mapping language.
  Ontop also supports the GeoSPARQL query language over
  PostgreSQL/PostGIS database. We use Ontop v5.0.2 in this experiment.
  
\item Apache Jena\footnote{\url{https://jena.apache.org/index.html}} is an open-source framework for Java that allows for reading, writing, and querying RDF graphs. Apache Jena Fuseki\footnote{\url{https://jena.apache.org/documentation/fuseki2/index.html}} is a sub-project of Jena which is a SPARQL server, combined with a UI for admin and query. TDB is a component of Jena for RDF storage and query which can be used as a high-performance RDF store. Unlike Ontop, Apache Jena Fuseki does not handle any SPARQL-to-SQL translation or virtualization but rather utilizes RDF triples as input. GeoSPARQL and spatial index support are available via the {jena-fuseki-geosparql} extension. Note that an RDF dataset needs to be ``wrapped'' as a GeoSPARQL dataset since the default Apache Jena Fuseki installation does not provide support for GeoSPARQL query functionalities.  We use Apache Jena v4.8.0 in this experiment.
  
\item  GraphDB\footnote{\url{https://graphdb.ontotext.com/}} is an RDF store developed by Ontotext, which supports SPARQL 1.1. OWL reasoning and is compliant with W3C Standards. It is a materialization-based system in a similar sense to Apache Jena, but although commercial, it does offer a free limited version. For the purpose of our experiments, we use GraphDB 10.2.2, which supports all SPARQL 1.1 and GeoSPARQL functionalities.
\end{itemize}
\begin{figure}[h!]
  \vspace{-0.5em} \centering
  \includegraphics[width=\textwidth]{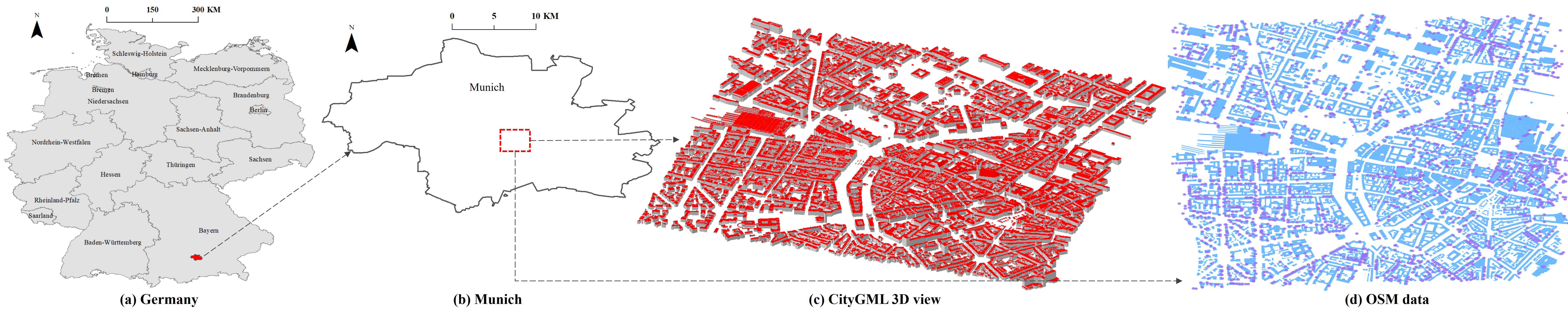}
  \caption{Study area.
  }
  \label{applied_data}
\end{figure}

\noindent\textit{Geographic Area of Interest.}
The experiments are carried out in the central area of Munich, Germany (\autoref{applied_data} (a) and (b)), which is a construction-dense area and can provide an adequate gauge of query performance. 
\autoref{applied_data}  (c) and (d) depict the test datasets in the area of interest from CityGML and  OSM respectively.


\subsection{Expressiveness test}

Geospatial queries are used in many application scenarios, e.g., urban planning or management, disaster management, tourism, and energy (solar panels). In order to test the expressiveness of GeoSPARQL queries that can be formulated over the KGs constructed from CityGML and OSM data in this work, we have collected 10 queries from domain experts and tried to formalize them as GeoSPARQL queries. 
These queries encompass not only conventional question types about 3D buildings but also those designed to address pragmatic real-world demands.

\subsubsection{Queries}

Queries Q1--Q5 represent basic information needs for 3D buildings:

\begin{itemize}[-]
\item Q1: Find the addresses of buildings with height above 30 meters
\item Q2: Find buildings with the address ``Stephansplatz''
\item Q3: Find 10 buildings that have the maximum number of roof surfaces
\item Q4: Find roof surfaces of buildings over 30 meters 
\item Q5: Find 3D geometries of buildings over 30 meters 
\end{itemize}

\smallskip

\noindent Queries 6--10 make use of both CityGML data and OSM data. 
Each case would encompass the possibility of being applied to a practical task involving the retrieval or analysis of real-world geospatial data. A summary of these queries are listed in Table~\ref{tab:prefixes}.

\smallskip

\noindent
If a researcher aims to perform a geometric analysis across OSM and CityGML data for different height ranges or building usage types, they might potentially have the following query: 
\begin{compactitem}[-]
\item Q6: Find CityGML ground surfaces and OSM building polygons for all residential buildings.
\end{compactitem}

\smallskip

\noindent For tourists who are seeking a hotel with a superior city view, they might inquire:
\begin{compactitem}[-]
\item Q7: Find hotels over 30 meters high
\end{compactitem}

\smallskip

\noindent In the context of emergency evacuation during a hurricane disaster, the inquiry might be posed as:
\begin{compactitem}[-]
\item Q8: Find residential buildings over 30m high.
\end{compactitem}

\smallskip

\noindent Various roof types such as gabled roofs and hip roofs possess the potential for installing photovoltaic panels. In the studied dataset, the roof type codes in CityGML follow the German cadastre information ALKIS codelists\footnote{\url{https://www.citygmlwiki.org/images/f/fd/RoofTypeTypeAdV-trans.xml}} for CityGML 2.0. This query could be framed as:
\begin{compactitem}[-]
\item Q9: Find residential buildings with non-flat roofs.
\end{compactitem}

\smallskip

\noindent Within the scope of urban renewal, individuals could inquire about the structures that could be affected and the conceivable expense or workload that might be earmarked for demolition:
\begin{compactitem}[-]
\item Q10: Find buildings along a certain road within 20m and calculate the total affected area in m$^2$.
\end{compactitem}

\begin{table*}[]
    \caption{Summary of the features used in Q6--Q10}
  \label{tab:queries}
\begin{center}
    \resizebox{\columnwidth}{!}{
\renewcommand{\arraystretch}{1.3}
\begin{tabular}{llll}
  \toprule
  & \textbf{CityGML} & \textbf{OSM} & \textbf{Filter} \\
  \hline
  Q6 & Building Geometry & Residential Building & Building Height \\
  Q7 & Building, Building Height & Hotel & Building Height \\
  Q8 & Building, Building Height & Residential Building & Building Height \\
  Q9 & Building, Roof Type & Residential Building & ALKIS RoofType \\
  Q10 & Building & Highway & Buffer and Intersection\\
  \bottomrule
\end{tabular}}
\end{center}
\end{table*}
\subsubsection{Results}
All of the ten queries could be successfully formulated via the SPARQL query language.
Below we provide the respective scripts for queries 9 and 10, while the remaining queries for our experiments can be found in \autoref{sec:appendixA}. The prefixes utilized are listed in \autoref{tab:prefixes} and refer to the base namespaces of the CityGML and LGD ontologies, where the remaining prefixes reference authoritative vocabularies such as RDFS and GeoSPARQL.\par

\begin{table}[]
\centering
  \caption{List of prefixes used for SPARQL queries.}
  \label{tab:prefixes}
\begin{tabular}{ll}
\toprule
Prefix & IRI Namespace \\
\midrule
\texttt{:} & \texttt{https://github.com/yuzzfeng/D2G2/citygml\#} \\
\texttt{bldg:} &  \texttt{http://www.opengis.net/citygml/building/2.0/} \\
\texttt{geo:} & \texttt{http://www.opengis.net/ont/geosparql\#}  \\
\texttt{rdfs:} & \texttt{http://www.w3.org/2000/01/rdf-schema\#} \\
\texttt{lgdo:} &  \texttt{http://linkedgeodata.org/ontology/} \\
\bottomrule
\end{tabular}
\end{table}

\smallskip
\noindent
\textit{Query 9} uses CityGML roof type codes in conjunction with their respective labels derived from ALKIS definitions and translated into English.

\begin{lstlisting}[language=SPARQL,basicstyle=\ttfamily\small]
SELECT ?citygmlGeom ?roofType
{
# Filter residential areas
?linkage a :Association_CityGML_OSM .
?linkage :matchOSM ?osmentity .
?linkage :matchCityGML/:mapSurface/bldg:bounds ?citygmlentity .
?osmlinkage a :Association_OSM_Class .
?osmlinkage :hasosmid ?osmentity .
?osmlinkage :hasosmclassid ?osmclassname .
VALUES ?residentialclasses { lgdo:Residential lgdo:ResidentialHome 
    lgdo:BuildingResidential lgdo:ApartmentBuilding lgdo:House } .
?osmclassname rdf:type ?residentialclasses .

# Filter non-flat roofs
?citygmlentity bldg:roofSurface/bldg:roofType ?roofType .
FILTER(?roofType != "flat roof") . 

# Retrieve geometry
?citygmlentity bldg:boundedBy ?citygmlsurface .
?citygmlsurface a bldg:RoofSurface .
?citygmlsurface geo:hasGeometry/geo:asWKT ?citygmlGeom .
}
\end{lstlisting}

\noindent
\textit{Query 10} utilizes two GeoSPARQL functions \texttt{geof:buffer} and \texttt{geof:sfIntersects} to both buffer and intersect geometries. It provides an example of how powerful geospatial functions can also be applied to linked data.

\begin{lstlisting}[language=SPARQL, basicstyle=\ttfamily\small]
SELECT ?citygmlGeom ?citygmlGeomAreaSqm
{
# Filter highways of interest
?osmlinkage a :Association_OSM_Class .
?osmlinkage :hasosmid ?osmentity .
?osmlinkage :hasosmclassid ?osmclassname .
VALUES ?highwayclasses { lgdo:SecondaryHighway lgdo:TertiaryHighway 
    lgdo:HighwayService lgdo:UnclassifiedHighway } .
?osmclassname rdf:type ?highwayclasses .
?osmentity rdfs:label ?street_name .
FILTER(CONTAINS(?street_name, "Elisenstraße")) .
?osmentity geo:hasGeometry/geo:asWKT ?osmGeom .

# Set impact area buffer
BIND(geof:buffer(?osmGeom, 20, uom:metre) AS ?impactArea) .
?citygmlentity bldg:boundedBy ?citygmlsurface .
?citygmlsurface a bldg:GroundSurface .
?citygmlsurface geo:hasGeometry/geo:asWKT ?citygmlGeom .

# Filter buildings within range of impact area
FILTER(geof:sfIntersects(?impactArea, ?citygmlGeom))
?citygmlsurface geo:hasGeometry/geo:hasMetricArea ?citygmlGeomAreaSqm .
}
\end{lstlisting}


\subsection{Performance test}

In some application scenarios, e.g. disaster management, expressiveness is not the sole aspect that matters, but the aspect of query execution time becomes critical. Given our dual virtual and materialized KG setup, we exploit this chance to assess whether our queries can be executed within a reasonable time but also how the KG setting might impact these results. The quantitative measures we analyze are database storage and query response time.

\subsubsection{Database size}

Although evaluating the storage requirements of each MKG and VKG solution was not a primary goal of our analysis, it can provide a useful indicator of scalability.
The CityGML and OSM data took up 501M of storage in PostgreSQL,  which includes not only the data but also geospatial indices. Ontop acting as a lightweight layer does not require additional storage, whereas Apache Jena and GraphDB store materialized triples generated by Ontop. Both Apache Jena and GraphDB use a 1.1G Turtle file of RDF triples to store the materialized triples. Moreover, for both these MKG solutions, we cannot create a geospatial index because an index cannot be added to a polyhedral surface or geometry collection respectively, rendering this figure a lower bound.





\subsubsection{Query response time}
\begin{figure}[hbt!]
    \vspace{-0.5em}
    \centering
    \includegraphics[width=\textwidth]{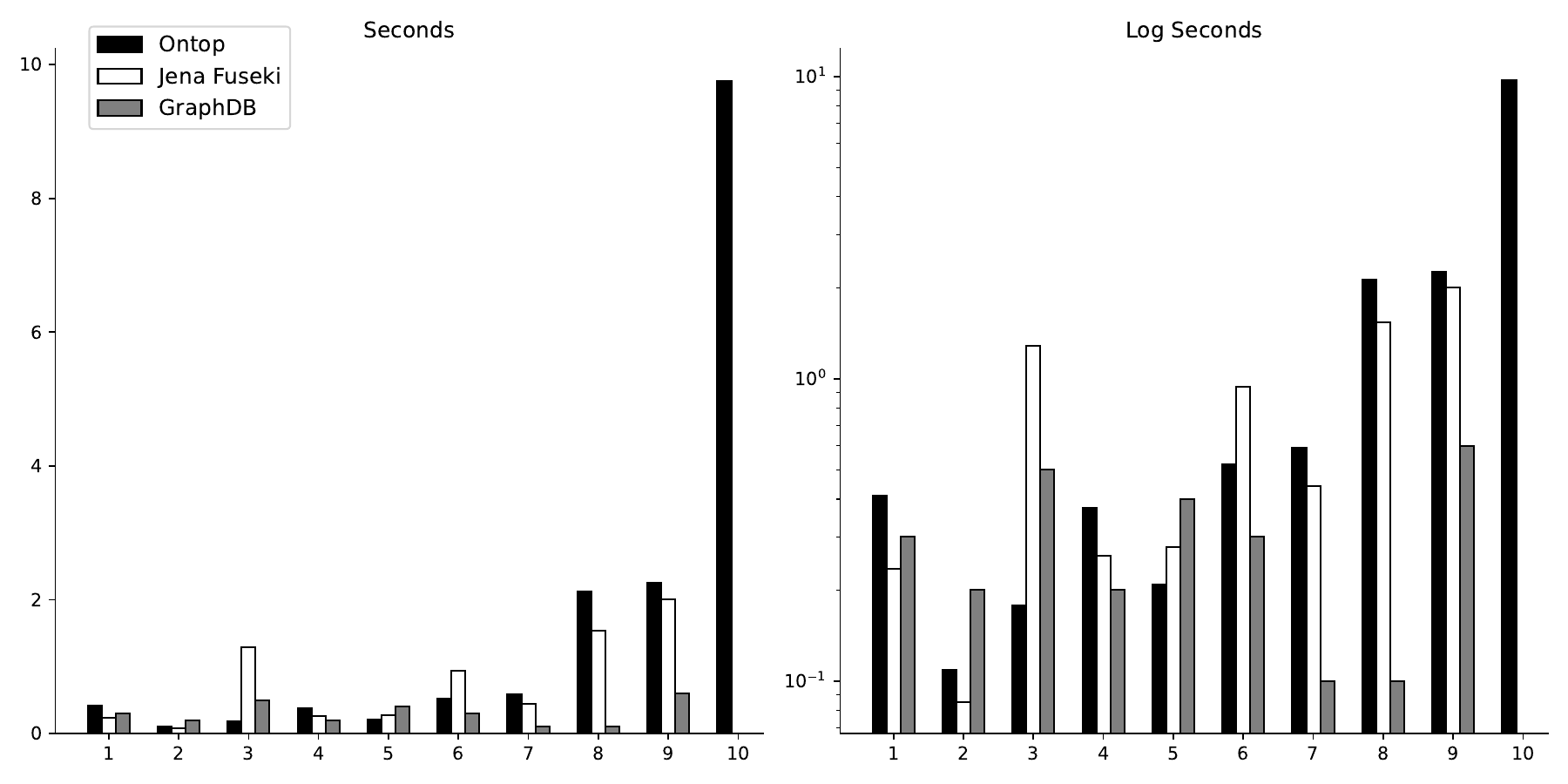}
    \caption{Query time performance by Q1--Q10}%
    \label{fig:timeperformancechart}%
\end{figure}

The results in terms of query response time in seconds are provided in~\autoref{tab:query_response_time} and visualized in~\autoref{fig:timeperformancechart}. Most of the queries can be evaluated by all the systems. All of the queries can be executed in under 10 seconds, with only one query exceeding 3 seconds. Given the relatively quick response time we deem overall performance satisfactory for a subjective domain practitioner for all queries.

\begin{table}[]
  \caption{Query response time}
  \label{tab:query_response_time}
\centering
\begin{tabular}{cccc}
\toprule
\textbf{Query} & \textbf{Ontop} & \textbf{Jena} & \textbf{GraphDB} \\
\midrule
Q1 & 0.411 & 0.235 & 0.3 \\
Q2 & 0.109 & 0.085 & 0.2 \\
Q3 & 0.178 & 1.285 & 0.5 \\
Q4 & 0.375 & 0.259 & 0.2 \\
Q5 & 0.209 & 0.278 & 0.4 \\
Q6 & 0.521 & 0.943 & 0.3 \\
Q7 & 0.592 & 0.441 & 0.1 \\
Q8 & 2.125 & 1.533 & 0.1 \\
Q9 & 2.259 & 2.007 & 0.6 \\
Q10 & 9.886 & NA & NA \\
\bottomrule
\end{tabular}
\end{table}

Due to limitations in their handling of non-simple features such as polyhedral surfaces for Apache Jena and geometry collections in the case of GraphDB, it was not possible to run Q10 which involves a geospatial function. Adding a geospatial index is not possible for either of these software solutions, polyhedral surfaces are part of the CityGML dataset whereas geometry collections of the OSM dataset. PostGIS is a more mature geospatial extension, and PostGIS version 3 does not exhibit any issues with indexing these geometry datatypes. A further drawback in the comparison for GraphDB is that the level of precision is limited to one decimal figure.\par

While there are variations across individual queries, performance is relatively similar for both the VKG and MKG solutions, no system outperforms across most queries. RDF stores tend to outperform Ontop for integrated queries mostly due to the large number of \texttt{UNION} clauses needed to assemble potential matching OSM data for multiple tag types i.e. node, way, relation and points of interest (e.g., ResidentialBuilding, House). GraphDB seems to display better performance compared to even Apache Jena for these large integrated scenarios.

\subsection{Qualitative comparison with pure relational databases}


In this section, we provide a qualitative comparison between SPARQL queries over the KGs, and equivalent SQL queries over relational database storing the original data. 
%
In general, since KGs represent a higher level of abstraction with the terminology used in the domain, it is easier to formulate queries in SPARQL, and the resulting SPARQL queries are more understandable.
Generating simple queries which rely solely on CityGML would be comparable in both SQL and SPARQL. A user would for example run a query retrieving building addresses by simply joining two tables from the SQL schema.
The task becomes considerably more difficult when an additional data source is integrated such as OSM. For example, we provide below the SQL translation of Query 10 formulated previously. We note that this SQL query is automatically generated by Ontop (slightly simplified for readability), but this would be rather close to what a human expert could produce. It would be extremely difficult and laborious for a human user to construct such a complex query.

\begin{lstlisting}[language=SQL, basicstyle=\ttfamily\small]
SELECT ST_ASTEXT(ST_TRANSFORM(v8."geometry1m27",4326)) AS "v1", 
ST_ASTEXT(ST_TRANSFORM(v8."geometry1m60",4326)) AS "v4"
FROM (SELECT DISTINCT v1."cityobject_id" AS "cityobject_id1m12", 
  v1."geometry" AS "geometry1m27", v2."geometry" AS "geometry1m60", 
  v1."id" AS "id1m11", v2."id" AS "id1m44", v3."osm_id" AS "osm_id1m451", 
  ST_ASTEXT(v4."geom") AS "v0", CAST(v5."building_id" AS TEXT) AS "v2", 
  CAST(v5."cityobject_id" AS TEXT) AS "v3"
FROM "surface_geometry" v1, "surface_geometry" v2, "public"."classes" v3, 
"public"."classes" v4, 
  (SELECT v1."building_id" AS "building_id", v2."cityobject_id" AS "cityobject_id"
  FROM "thematic_surface" v1 LEFT JOIN "surface_geometry" v2 
    ON v1."lod2_multi_surface_id" = v2."root_id") v5, 
  (SELECT v1."id" AS "id"
  FROM "cityobject" v1 LEFT JOIN "objectclass" v2 ON v1."objectclass_id"=v2."id"
  WHERE v2."classname"='BuildingGroundSurface') v6
WHERE (ST_INTERSECTS(ST_BUFFER(CAST(ST_ASTEXT(v4."geom") AS GEOGRAPHY),'20'),
CAST(ST_TRANSFORM(v1."geometry",4326) AS GEOGRAPHY)) 
AND v5."cityobject_id" = v1."cityobject_id"
AND v5."cityobject_id" = v6."id" 
AND v1."cityobject_id" = v2."cityobject_id" 
AND v3."osm_id" = v4."osm_id" 
AND ('W' = v3."osm_type" AND 'SecondaryHighway' = v3."class") 
AND 'W' = v4."osm_type")
) v8
\end{lstlisting}

\section{Discussion and Future Work}
\label{sec:discussion}
This paper presents a comprehensive framework to analyze 3D City data via KGs. It provides a methodology to integrate CityGML data with other geospatial data sources and utilize the resulting KG to answer user queries. The experiments confirm that the expressive queries can be formulated over the KGs, and can be efficiently evaluated with state-of-the-art KG systems.
While the obtained results are promising, below we discuss some limitations that arose while running the experiments, and possible future improvements.

\smallskip
\noindent\textit{Support for Complex Geometries.} Many KG systems have limitations with respect to the handling of more complex geometries such as polyhedral surfaces and geometry collections, which could not be parsed correctly and missed support of respective geospatial index. Specifically, this meant that our MKG approach (GraphDB and Apache Jena) suffered with respect to the execution of Q10 that involves computation with complex geometries.
Instead, the VKG system Ontop can leverage mature and well-established relational spatial databases such as PostGIS, and thus avoided such issues.

\smallskip
\noindent\textit{CityGML ontology.} While the  CityGML ontology  developed by University of Geneva was selected as the most renowned choice available, its future adoption as the central ontology for 3D building analysis displays conspicuous limitations. The most evident limitation is caused by its inconsistencies in the duplication of the definition of data and object properties which were detailed in~\autoref{sec:architecture} and in~\cite{CHADZYNSKI2021100106}.

\smallskip
\noindent\textit{CityGML 3.0.}
The CityGML ontology in this work supports only the CityGML specification up to version 2. The latest version of CityGML, version 3, not only introduces new concepts such as  time-dependent features but also revises the existing specification \textit{e.g.,} dropping LoD4~\cite{citygml3}. A new ontology will need to be designed or selected for future semantic analysis of CityGML data. Moreover, the version of 3DCityDB system for CityGML 3.0 is still under development at the time of writing. In order to support CityGML 3.0 in our framework, the VKG mapping also need to be revised with respect to the new ontology and 3DCityDB schema.


\smallskip
\noindent\textit{CityGML heterogeneity.} The CityGML data produced by and with the specifications of the government of Bavaria, Germany was used for this analysis. However, during the study, we found that different countries might have different standards for encoding their CityGML data which makes issues such as SRID differences arise. For example, Estonia also provides a geometry element with its address, and it links each building not to the corresponding 3D solid but to individual surface geometries (failing to provide any solid geometries)\footnote{\url{https://geoportaal.maaamet.ee/eng/Download-3D-data-p837.html}}. Hence, the design of VKG mappings to link the ontology with the same 3DCityDB physical storage is not guaranteed to be robust for analyses across countries. The mapping should be tested for robustness by repeating these experiments with datasets from as many countries as possible to strive to reach a unified design.

\smallskip
\noindent\textit{CityGML data paucity.} Although a significant degree of expressiveness was successfully tested by leveraging the data on CityGML buildings, there is data paucity for both higher levels of detail such as LOD3, and other non-building items such as vegetation, waterbodies, bridges, etc. 
 The lack of this data makes a significant portion of the 3DCityDB SQL schema redundant for almost all publicly available CityGML datasets. 

\smallskip
\noindent\textit{Integrating further data source.} Our paradigm and evaluation sought to measure CityGML and OSM data integration. While OSM is a popular source of geospatial data,  our analysis might not be generalized to more unique geospatial domains. Expressiveness and the respective overall performance might diminish with the introduction of additional types and combinations of geospatial data. Tasks such as the complexity of matching objects might correspondingly become more complex and have an impact on both ontology integration and query design. These possible risks can be only addressed through further experimental research.


\section*{Acknowledgements}
This research has been partially supported by German Research Foundation (DFG) and the Autonomous Province of Bolzano-Bozen through its Joint Project - Dense and Deep Geographic Virtual Knowledge Graphs for Visual Analysis - D2G2.

\begin{figure}[hbt!]
  \vspace{-0.5em}
  \includegraphics[width=0.2\textwidth]{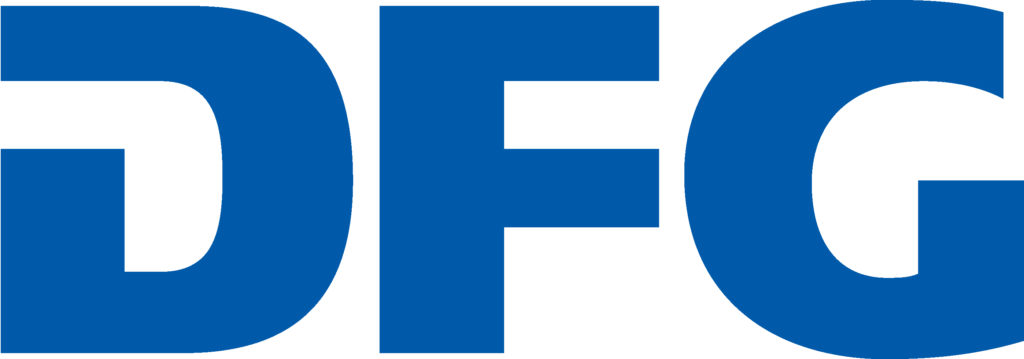}
    \includegraphics[width=0.3\textwidth]{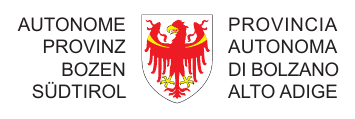}
\end{figure}    

\appendix
\section{Appendix A}
\label{sec:appendixA}
\paragraph{Query 1} Find the addresses of buildings with height above 30 meters.

\begin{lstlisting}[language=SPARQL,basicstyle=\ttfamily\small]
SELECT ?address_label
{
?building bldg:address ?address_id .
?address_id rdfs:label ?address_label .
?building bldg:measuredHeight ?buildingHeight .
FILTER(?buildingHeight > 30) .
}
\end{lstlisting}

\paragraph{Query 2} Find buildings with the address “Stephansplatz”

\begin{lstlisting}[language=SPARQL,basicstyle=\ttfamily\small]
SELECT ?building ?address_label
{
?building bldg:address ?address_id .
?address_id rdfs:label ?address_label .
FILTER(CONTAINS(?address_label, "Stephansplatz")) .
}
\end{lstlisting}

\paragraph{Query 3} Find 10 buildings that have the maximum number of roof surfaces

\begin{lstlisting}[language=SPARQL,basicstyle=\ttfamily\small]
SELECT ?building (COUNT(?surface) AS ?totalsurface)
{
?building a bldg:Building .
?building bldg:boundedBy ?surface .
?surface a bldg:RoofSurface .
}
GROUP BY ?building
ORDER BY DESC(?totalsurface)
LIMIT 10
\end{lstlisting}

\paragraph{Query 4} Find roof surfaces of buildings over 30 meters

\begin{lstlisting}[language=SPARQL,basicstyle=\ttfamily\small]
SELECT ?citygmlGeom
{
?citygmlentity bldg:measuredHeight ?citygmlBuildingHeight .
FILTER(?citygmlBuildingHeight > 30) .
?citygmlentity bldg:boundedBy ?citygmlsurface .
?citygmlsurface a bldg:RoofSurface .
?citygmlsurface geo:hasGeometry/geo:asWKT ?citygmlGeom .
BIND("chlorophyll,0.5" AS ?citygmlGeomColor) # Green
}
\end{lstlisting}

\paragraph{Query 5} Find 3D geometries of buildings over 30 meters

\begin{lstlisting}[language=SPARQL,basicstyle=\ttfamily\small]
SELECT ?citygmlGeom
{
?citygmlentity bldg:measuredHeight ?citygmlBuildingHeight .
FILTER(?citygmlBuildingHeight > 30) .
?citygmlentity bldg:lod2Solid ?solid .
?solid geo:asWKT ?citygmlGeom .
BIND("chlorophyll,0.5" AS ?citygmlGeomColor) # Green
}
\end{lstlisting}

\paragraph{Query 6} Find CityGML ground surfaces and OSM building polygons for all residential
buildings

\begin{lstlisting}[language=SPARQL,basicstyle=\ttfamily\small]
SELECT ?citygmlGeom ?osmGeom
{
?linkage a :Association_CityGML_OSM .
?linkage :matchOSM ?osmentity .
?linkage :matchCityGML/:mapSurface/bldg:bounds ?citygmlentity .
?citygmlentity bldg:measuredHeight ?citygmlBuildingHeight .
FILTER(?citygmlBuildingHeight > 30) .
?citygmlentity bldg:boundedBy ?citygmlsurface .
?citygmlsurface a bldg:GroundSurface .
?citygmlsurface geo:hasGeometry/geo:asWKT ?citygmlGeom .
BIND("chlorophyll,0.5" AS ?citygmlGeomColor) # Green
?osmentity geo:hasGeometry/geo:asWKT ?osmGeom .
BIND("jet,0.8" AS ?osmGeomColor) # Red
}
\end{lstlisting}

\paragraph{Query 7} Find hotels over 30 meters high

\begin{lstlisting}[language=SPARQL,basicstyle=\ttfamily\small]
SELECT ?citygmlentity ?buildingHeight ?hotelname ?citygmlGeom
{
?linkage a :Association_CityGML_OSM .
?linkage :matchOSM ?osmentity .
?linkage :matchCityGML/:mapSurface/bldg:bounds ?citygmlentity .
?osmlinkage a :Association_OSM_Class .
?osmlinkage :hasosmid ?osmentity .
?osmlinkage :hasosmclassid ?osmclassname .
?osmclassname a lgdo:Hotel .
    OPTIONAL { ?osmentity rdfs:label ?hotelname .}
?citygmlentity bldg:measuredHeight ?buildingHeight .
FILTER(?buildingHeight > 30) .
?citygmlentity bldg:lod2Solid ?solid .
?solid geo:asWKT ?citygmlGeom .
}
\end{lstlisting}

\paragraph{Query 8} Find residential buildings over 30m high

\begin{lstlisting}[language=SPARQL,basicstyle=\ttfamily\small]
SELECT ?citygmlentity ?citygmlGeom
{
?linkage a :Association_CityGML_OSM .
?linkage :matchOSM ?osmentity .
?linkage :matchCityGML/:mapSurface/bldg:bounds ?citygmlentity .
?osmlinkage a :Association_OSM_Class .
?osmlinkage :hasosmid ?osmentity .
?osmlinkage :hasosmclassid ?osmclassname .
# Known Residential categorization
VALUES ?residentialclasses { lgdo:Residential lgdo:ResidentialHome
lgdo:BuildingResidential lgdo:ApartmentBuilding lgdo:House } .
?osmclassname rdf:type ?residentialclasses .
?citygmlentity bldg:measuredHeight ?buildingHeight .
FILTER(?buildingHeight > 30) .
?citygmlentity bldg:boundedBy ?citygmlsurface .
?citygmlsurface a bldg:GroundSurface .
?citygmlsurface geo:hasGeometry/geo:asWKT ?citygmlGeom .
BIND("chlorophyll,0.5" AS ?citygmlGeomColor) # Green
}
\end{lstlisting}


\printbibliography

\end{document}